\pdfoutput=1
\RequirePackage{fix-cm}
\documentclass{svjour3}                     
\usepackage{graphicx}
\usepackage{amsmath,amsfonts,amssymb}
\usepackage{algorithm}
\usepackage{booktabs}
\usepackage{bm}
\usepackage{hyperref}
\newcommand{\dd}{\mathrm{d}}
\newcommand{\Var}{\operatorname{Var}}
\newcommand{\E}{\mathbb{E}}
\newcommand{\N}{\mathcal{N}}

\newcommand{\bgamma}{{\bm{\gamma}}}
\newcommand{\brho}{\bm{\rho}}

\newcommand{\beps}{\bm{\varepsilon}}
\newcommand{\bxi}{\bm{\xi}}

\newcommand{\B}{{B}}
\newcommand{\bB}{\bm{b}}

\newcommand{\bc}{\bm{c}}
\newcommand{\bd}{\bm{d}}
\newcommand{\be}{\bm{e}}
\newcommand{\bx}{\bm{x}}
\newcommand{\by}{\bm{y}}

\usepackage{color}
\newcommand{\cf}[1]{#1}

\newtheorem{theo}{Theorem}
\newtheorem{coro}{Corollary}
\newtheorem{appr}{Approximation}

\begin{document}

\title{Randomized Reduced Forward Models for Efficient Metropolis--Hastings MCMC, 
       with Application to Subsurface Fluid Flow and Capacitance Tomography
\thanks{This is a pre-print of an article submitted to GEM - International Journal on Geomathematics. 
}
}

\titlerunning{Randomized Reduced Models for MCMC}        

\author{Colin Fox*         \and
        Tiangang Cui      \and
        Markus Neumayer
}

\authorrunning{Fox Cui \& Neumayer} 

\institute{C. Fox \at
              Department of Physics, University of Otago, Dunedin, New Zealand \\
              \email{colin.fox@otago.ac.nz}     (corresponding author)      
           \and
           T. Cui \at
           School of Mathematics, Monash University, Melbourne, Australia \\
           \email{tiangang.cui@monash.edu} 
            \and
           M. Neumayer \at
           Institute of Electrical Measurement and Sensor Systems, Graz University of Technology, Graz, Austria \\
           \email{neumayer@tugraz.at}
}

\date{Received: date / Accepted: date}

\maketitle

\begin{abstract}
Bayesian modelling and computational inference by Markov chain Monte Carlo (MCMC) is a principled 
framework for large-scale uncertainty quantification, though is limited in practice by computational cost when implemented in the simplest form that requires simulating an accurate computer model at each iteration of the MCMC. 
The delayed acceptance Metropolis--Hastings MCMC leverages a reduced model for the forward map to lower the compute cost per iteration, though necessarily reduces statistical efficiency that can, without care, lead to no reduction in the computational cost of computing estimates to a desired accuracy. 
Randomizing the reduced model for the forward map can dramatically improve computational efficiency, by maintaining the low cost per iteration but also avoiding appreciable loss of statistical efficiency. 
Randomized maps are constructed by \textit{a posteriori} adaptive tuning of a randomized and locally-corrected deterministic reduced model.  
Equivalently, the approximated posterior distribution may be viewed as induced by a modified likelihood function for use with the reduced map, with parameters tuned to optimize the quality of the approximation to the correct posterior distribution.
Conditions for adaptive MCMC algorithms allow practical approximations and algorithms that have guaranteed ergodicity for the target distribution.
Good statistical and computational efficiencies are demonstrated in examples of calibration of large-scale numerical models of geothermal reservoirs and electrical capacitance tomography. 
\end{abstract}

\section{Introduction: Background and Context}
\label{sec:intro}
 
Characterizing subsurface properties in geosciences and performing non-invasive imaging for industrial process monitoring are typical examples of inverse problems. In this paper we present computational Bayesian methods developed for such inverse problems, and present computed examples for two cases: calibrating numerical models of geothermal reservoirs, and performing electrical capacitance tomography (ECT) inside a pipe. In each case, the desired physical properties are inferred from indirect observations made on the system and the associated inverse problem has several fundamental difficulties: data are sparsely measured and corrupted by noise, the forward model requires solving a partial differential equation (PDE) and only has a limited range of accuracy in representing the underlying system, and the parameters of interest are spatially distributed and highly heterogeneous.

Consider the case where the physical measurement system is simulated by a forward model $F(\cdot)$.
At the true, unknown parameters $\bx$, 
\[ {\bd} = F({\bx})\]
is the  noise-free data. 
Practical measurements $\tilde{\bd}$ are a noisy version of  ${\bd}$, being subject to measurement errors and model error.

The inverse problem is to estimate the unknown $\bx$ from measurements $\tilde{\bd}$. The related prediction problem is to infer properties of the physical system such as future, unobserved data. 
All physical forward maps $F$ effectively have finite rank~\cite {BirmanSolomyak1977}, hence the model output occupies some low dimensional manifold in data space which causes the inverse problem to be ill-posed~\cite{hadamard,KaipioSomersalo2004}, and correlations between estimated parameters to be extremely high. 
These properties make the traditional, deterministic solution to the inverse problem very sensitive to measurement error and model error.

Uncertainty in measured data $\tilde{\bd}$, in the model $F(\cdot)$, and in possible values for $\bx$, leads to uncertainty in estimates of $\bx$ and in subsequent predictions. 
What then is the range of permissible values of ${\bx}$, or of predicted properties, that is, what is the implied distribution over resulting estimates? Bayesian modelling and inference provides a principled route to quantifying these estimates and uncertainties.

\begin{figure}[h]
   \centerline{\includegraphics[scale=1]{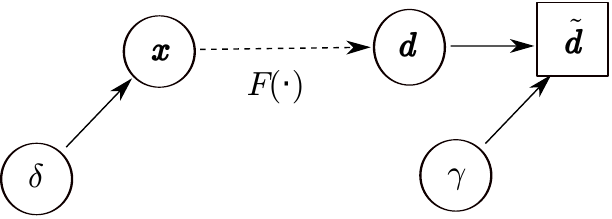}}
   \caption{A DAG for a practical inverse problem, showing the exact forward map $F$. 
   Shown are the unknown latent field ${\bx}$, measured data $\tilde{\bd}$, and also hyperparameters $\gamma$ and $\delta$ that represent uncertainties in the observation process and in stochastic modelling of ${\bx}$, respectively. Random variables are depicted as nodes; circular nodes depict unknown variables while square nodes indicate observed values. Stochastic dependencies are shown as solid directed lines (for example, $\tilde{\bd}$ is a measured random variable with distribution parameterized by unknown noise-free data $\bd$ and unknown hyperparameter $\gamma$), while the deterministic forward map $F$ is depicted by a dashed line signifying that it is redundant when specifying conditional dependencies (though not functional forms).\label{fig:dagip}}
\end{figure}

The Bayesian formulation of an inverse problem requires modelling all functional and conditional dependencies between variables, and assigning probability distributions to each source of error, or uncertainty. Functional and conditional dependencies are conveniently displayed as a Bayesian network, a.k.a., directed acyclic graph (DAG). Figure~\ref{fig:dagip} shows a DAG for a practical inverse problem.  
  
The forward probability problem is to determine the distribution over possible data  $\tilde{\bd}$, by following the conditional dependencies in a forward direction, while the inverse probability, or `Bayesian', problem is to determine the distribution over unknown random variables given measured data $\tilde{\bd}$, effectively following the conditional dependencies in a reverse direction.
By Bayes rule, the (unnormalized) posterior distribution over unknown random variables, conditioned on measured data $\tilde{\bd}$, is conveniently given by the product of distributions appearing in the hierarchical model,
\begin{equation}
 \pi_\text{post}({\bx}, \gamma, \delta | \tilde{\bd}) \propto L(\tilde{\bd} | {\bx}, \gamma)  \pi_\text{prior}({\bx} | \delta) \pi_\Delta(\delta)  \pi_\Gamma(\gamma) .
 \label{eq:bayes}
\end{equation}
Here $L(\tilde{\bd} | {\bx}, \gamma)$ denotes the conditional distribution over measured data conditioned on $\bx$ and hyperparameter $\gamma$ that is called the likelihood function when viewed as a function of $\bx$ and hyperparameters, $\pi_\text{prior}({\bx} | \delta)$ is the prior distribution over $\bx$, while $\pi_\Delta(\delta)$ and $\pi_\Gamma(\gamma)$ are hyperprior distributions over hyperparameters $\delta$ and $\gamma$, respectively. 

Measured data is commonly assumed to be related to noise-free data by the additive error model
\begin{equation}
\tilde{\bd} = F({\bx}) + \be,
\label{eq:cali}
\end{equation}
where the random vector $\be$ captures the measurement noise and other uncertainties such as model error. 
When $\be$ follows a zero mean multivariate Gaussian distribution~\cite{model_higdon2003} the resulting likelihood function has the form 
\begin{equation}
L(\tilde{\bd} | {\bx}, \Sigma_{\be}) \propto \exp \left[ -\frac{1}{2}  \{ F({\bx})-\tilde{\bd} \} ^T \mathbf{\Sigma}_{\be}^{-1} \{ F({\bx})-\tilde{\bd}\} \right],
\label{eq:llkd1}
\end{equation}
where the hyperparameter $\gamma = \mathbf{\Sigma}_{\be}$ is the covariance matrix of the noise vector $\be$, 
with uncertainty in the covariance being modelled by the (hyper)prior distribution $\pi_\Gamma(\gamma)$.

The contribution of model error to the noise vector $\be$ is usually non-negligible. This may be caused by discretization error in the computer implementation of the mathematical forward model and/or wrong assumptions in the mathematical model. 
We follow~\cite{model_higdon2003} who observe that it may not be possible to separate the measurement noise and model error when only a single set of data is available. 
Thus, it is necessary to incorporate the modeller's judgments about the appropriate size and nature of the noise term $\be$.

High correlations between the primary parameter $\bx$ and hyperparameters $\gamma$ and $\delta$ are typical, and introduce significant computational difficulties \cite{gmrf_rue}, though can be circumvented by sampling from the marginal posterior distribution over hyperparameters~\cite{FoxNorton2016,NortonChristenFox2018}. We do not consider that complexity here, and set the value of hyperparameters based on expert opinion and field measurements. This yields the posterior distribution
\begin{equation}
\pi_\text{post}({\bx}|\tilde{\bd}) \propto \exp\left[-\frac{1}{2}  \{ F({\bx})-{\tilde{\bd}} \}^T \mathbf{\Sigma}_{\be}^{-1} \{ F({\bx})-{\tilde{\bd}} \}\right] \pi_\text{prior}({\bx})
\label{eq:post}
\end{equation}
that is used throughout the remainder of this paper.

Evaluating the likelihood function for a particular ${\bx}$ requires simulating the forward model $F(\bx)$, which is a computationally expensive simulation of the physical system.
Computational efficiency can be significantly improved by exploiting reduced models; indeed, one could say that reduced models are \emph{mandatory} in inverse problems since discrete computer models typically approximate a function-space mathematical model. 

While recent developments in the Bayesian formulation of inverse problems on function spaces~\cite{lie2018random} have emphasized that a consistent discretization of the forward map (plus other conditions) leads to the computed posterior distribution converging to the function-space posterior distribution in the limit of refined discretization (and exact arithmetic), the concern of practical computing for large-scale inverse problems is always at the other end of the computational scale, i.e., finding the cheapest possible, perhaps crude, computational approximation to the forward map that can give sufficiently accurate estimates of quantities of interest. 
The focus of this paper is on improving such reduced models by developing random corrections that significantly increase the accuracy of estimates, at no significant increase in computational cost. We thereby change the cost/accuracy trade-off to allow even cheaper approximations to be used to achieve a desired accuracy in estimates.

The starting point for the methods developed here is a deterministic reduced model $F^*$ built using one of the many standard methods, such as:
\begin{itemize}
 \item grid coarsening, e.g.,~\cite{upscaling_christie_2001,KaipioSomersalo2007,Efendiev}, or a nested hierarchy of grid discretizations~\cite{DodwellKetelsenScheichlTeckentrup2015},
 \item global linearization of the forward map, i.e., the Born approximation~\cite{BerteroBoccacci1998},
 \item local linearization of the forward model, e.g.~\cite{ChristenFox2005},
 \item projection-based methods, e.g.~\cite{mor_BGW_2015,MOR_DEIM_reservoir,mor_LWG_2015,mor_CMW_2015,mor_APL_2016},
 \item anything else you can think of, including \textit{ad hoc} calibration of very coarse discretizations, e.g.~\cite{NeumayerPhD}; see Section~\ref{sec:ect}.
\end{itemize}
The delayed acceptance (DA) algorithm~\cite{ChristenFox2005} (see Section~\ref{sec:da}) is a compound Metropolis--Hastings algorithm that utilizes any reduced model to lower the computational cost per iteration, while correctly targeting the posterior distribution\footnote{In some sense DA is \emph{the most general} MH method for utilizing a single reduced model while maintaining the target distribution~\cite{BanterleGrazianLeeRobert2019}.}. We build on DA in this paper. 

\cf{DA, and all algorithms presented in this paper, target the correct posterior distribution by occasionally computing the exact forward map $F$. This paper addresses the practical question of how crude, and hence cheap, can be the reduced forward map $F^*$, and how few evaluations of the exact, expensive forward map can we get away with to achieve maximal computational efficiency. Of course the ideal situation is that computing $F^*$ is free and is sufficiently accurate that no evaluations of $F$ are required; we get remarkably close to that ideal in the computed examples in Section~\ref{sec:compex}, using an adaptive DA algorithm that corrects the na\"ive approximate posterior distribution defined by $F^*$.}

When a fixed, i.e., state-independent, approximation $F^*$ is used, DA simplifies to the surrogate transition method~\cite{mcmc_liu},  rediscovered as preconditioned or two-step MH~\cite{Efendiev}. This is Approximation~\ref{appr:coarse} in Section~\ref{sec:approx}. The approximate posterior distribution is given by analyzing the DAG in Figure~\ref{fig:dagip} with $F$ replaced by $F^*$. Using a fixed deterministic reduced model within DA/surrogate transition does not actually improve computational efficiency in any computed examples of geophysical inverse problems, that we are aware of; see Section~\ref{sec:geo}.

One observation that we want readers to take from this paper is that the approximation to the posterior distribution and accuracy of calculated estimates can be significantly improved by \emph{randomizing} the deterministic reduced model. The primary randomizing used throughout this paper is to add a random variable to the output of the deterministic reduced model, though other randomizing can be more effective in some settings; see, e.g.~\cite{OliverMoser2011}. That is, we model noise-free data as
\begin{equation}
  {\bd} = F^*({\bx}) + \bB
  \label{eq:addrand}
\end{equation}
where $\bB$ is independently drawn from some distribution to be determined, usually Gaussian. Adding independent random variables, as in Eqn~\eqref{eq:addrand}, corresponds to convolution of distributions. Our intuition is that the shift and smudging-out of the approximate posterior distribution by Eqn~\eqref{eq:addrand} leads to a better alignment of the support of the true and resulting approximate distributions. 

\begin{figure}[h]
   \centerline{\includegraphics[scale=1]{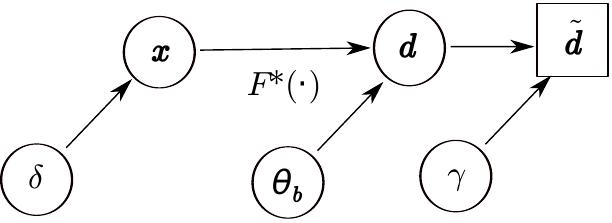}}
   \caption{DAG showing the randomized approximate forward map $F^*$, with the conditional distribution over noise-free data depending on variable $\theta_{\bB}$.\label{fig:dagstar}}
\end{figure}

In particular, we replace the DAG in Fig.~\ref{fig:dagip} by the DAG in Fig.~\ref{fig:dagstar} for approximating the posterior distribution, with the conditional distribution over noise-free data depending on parameters $\theta_{\bB}$ to be determined. 
This leads to the approximation of the expression in~\eqref{eq:bayes} by
\begin{equation}
 \pi^*({\bx}, \gamma, \delta | \tilde{d}) \propto L^*(\tilde{\bd} | {\bx}, \gamma, \theta_{\bB})  \pi_\text{prior}({\bx} | \delta) \pi_\Delta(\delta) \pi_\Gamma(\gamma) .
 \label{eq:bayesstar}
\end{equation}
A fully Bayesian analysis might place a hyperprior distribution over  $\theta_{\bB}$ and seek a posterior distribution; we find it sufficient to evaluate a best value.

Note that randomizing the forward map does not mean that evaluating the approximated likelihood requires generating a random vector, such as $\bB$ in the additive example above. Rather, the randomizing distribution is chosen to have a known form so that the (unknown) random term may be marginalized over, just as the (additive) noisy observation model~\eqref{eq:cali} leads to the likelihood function~\eqref{eq:llkd1} that requires simulating the exact forward map $F$, only. In particular, evaluating the approximate likelihood function induced by a randomized reduced forward model requires simulating the deterministic reduced map, only. See Approximations~\ref{appr:aemsi} and \ref{appr:aemsd} in Section~\ref{sec:approx}. Modification of the likelihood function is expanded upon in Section~\ref{sec:view}.

The observation that randomizing a reduced model can significantly improve the approximation of the posterior distribution may come as a surprise to numerical analysts who typically derive the deterministic reduced model according to some measure of `best'. However, we note that notion of `best' is applied to the forward map and not the posterior distribution. On the other hand, the improvement should be no surprise since randomized models include the deterministic model as a special case, so are not necessarily worse. Perhaps the real surprise is that randomizing can give an improvement in computational efficiency by more than a factor of five, in the large-scale examples we compute, for very little extra coding effort; see Section~\ref{sec:1d}.

\subsection{Correcting the Error Introduced by an Approximate Forward Map} 

The idea of randomizing a deterministic reduced model to improve the approximation has been used before. \cf{Here, we wish to provide a new interpretation that leads to a greater range of computational efficiencies. To explain this, we first briefly review existing stochastic models for the error induced by using a reduced model.

\subsubsection{A Brief History of Model Error}

Kennedy and O'Hagan~\cite{KennedyO'Hagan2000,KennedyO'Hagan2001} developed a Bayesian framework for \emph{calibrating} complex computer codes designed to simulate a scalar physical observation process $F(\cdot)$, noting that all computer codes approximate the true process. Their concern was to infer $F(\bx)$ at some point $\bx$ based on a fixed set of evaluations of $F$ and approximate model $F^*$, or multiple approximate models, at a set of points that may not include $\bx$. They introduced the use of a Gaussian process (GP) to model the function $F$ and approximation(s) $F^*$ based on the assumption of smoothness of functions, and the desirability of a non-parametric stochastic model for smooth functions; a GP is a common and flexible such model. They modeled the relationship between approximate and exact maps by $F(\cdot)=\rho F^*(\cdot)+\bB$ where $\rho$ is a kind of regression parameter and $\bB$ is distributed as a GP. Fitting of parameters in this stochastic model was performed by posterior inference.

For inverse problems defined over a high-dimensional space, a GP model typically does not provide computationally efficient and accurate function approximation. For example, even storing the mean function when representing $F$ is tantamount to storing sufficient output of the function  $F(\cdot)$ to be able to interpolate an accurate, `best' approximation~\cite{KennedyO'Hagan2000}. Instead, it is typically far cheaper, at a given accuracy, to recompute an approximate function $F^*(\bx)$ built using one of the many well-developed methods, listed above, that are designed to capture the basic structure of $F(\cdot)$ while being computationally efficient. The error introduced by this approximation, $F(\bx)-F^*(\bx)$, is then modelled by a Gaussian distribution that is a simplified case of the GP.

In spatial statistics, this often results in a `PDE-based' Markov random field (MRF) formulation; see~\cite[Section 6.4]{higdon2006} and references therein. For example, when evaluating $F$ requires simulating a system of ordinary differential equations (ODEs) using, say, the Euler step $F(\bx,t+\dd t)= F(\bx,t)+ \dd t f(\bx)$ for some known $f$, and the approximation uses a large time step $\Delta t$ and approximation $f^*$ and the step-wise approximation error $\bB$ is modelled as Gaussian, the resulting stochastic system of ODEs $F^*(\bx,t+\Delta t)= F^*(\bx,t)+ \Delta t f^*(\bx) +\bB$~\cite{higdon2006} defines a stochastic MRF approximation to $F$. Typically $f^*$ is linear and the true distribution over initial conditions is used, as in filtering applications, giving a Gaussian model for $F(\cdot)-F^*(\cdot)$.

Kaipio and Somersalo~\cite{KaipioSomersalo2004,KaipioSomersalo2007} also used a Gaussian error model when using approximative forward maps in inverse problems, writing $F(\cdot)-F^*(\cdot)=\B(\cdot)$ and then modelling unknown $\B(\cdot)$ by random variable $\bB$ that is independent of $\bx$ and Gaussian, i.e., ${\bB} \sim N({\boldsymbol \mu}_{\bB}, \mathbf{\Sigma}_{\bB})$. Substituting this, or equivalently Eqn~\eqref{eq:addrand}, into the observation model Eqn~\eqref{eq:cali} gives 
\begin{equation}
\tilde{\bd} = F^*({\bx}) + \be + \bB,
\label{eq:eem}
\end{equation}
that they called the enhanced error model (EEM)~\cite{KaipioSomersalo2004,KaipioSomersalo2007}.
These works only report estimates that maximize the posterior density conditioned on fixed hyperparameters for Gaussian likelihood functions, commonly known as regularized inversion or ridge regression. These conditional posterior modes correspond to posterior expectations only in the linear-Gaussian case and when uncertainty in hyperparameters is neglected\footnote{Uncertainty in hyperparameters causes the Bayesian posterior distribution to be non-Gaussian, and no conditional posterior mode as evaluated in~\cite{KaipioSomersalo2004,KaipioSomersalo2007} is a good approximation to the Bayesian posterior mode or mean, even for linear forward maps and Gaussian stochastic models~\cite{SimpsonLindgrenRue2012,FoxNorton2016}.}. That linear-Gaussian case was analyzed in~\cite{KaipioSomersalo2007} 
with the EEM giving improved approximated conditional posterior modes in a computed example, over a range of noise to discretization ratios\footnote{The ratio of noise standard deviation to length scale relative to discretization level is the regularization parameter in regularized inversion\cite{BardsleyRTO,FoxNorton2016}.}; see~\cite[Fig. 5]{KaipioSomersalo2007} and  \cite[Fig. 5.22]{KaipioSomersalo2004}. 
For non-linear forward map $F$, for which the matrix-vector calculations of the linear-Gaussian case are not sufficient, Kaipio and Somersalo used an off-line calculation to tune the EEM by using sample-based estimates of ${\boldsymbol \mu}_{\bB}$ and $\mathbf{\Sigma}_{\bB}$ from evaluations of $\B(\bx)=F(\bx)-F^*(\bx)$ with $\bx$ drawn from the \emph{prior} distribution over $\bx$~\cite[Section 7.6]{KaipioSomersalo2004} (this is Approximation~\ref{appr:aemsi} with \textit{a priori} EEM, in Section~\ref{sec:approx}). A computed example of EIT showed that this \textit{a priori} EEM produces approximated conditional posterior modes that better approximate the true conditional posterior mode compared to no EEM; see~\cite[Fig. 7.31]{KaipioSomersalo2004}.

Cui, Fox and O’Sullivan~\cite{wrr_cfo_2011,CuiFoxO'Sullivan2019} reinterpreted the empirical results in~\cite[Fig. 7.31]{KaipioSomersalo2004}, arguing that it seemed likely that the EEM  also increases the quality of the approximation to the Bayesian posterior distribution. However, \textit{a priori} tuning of the EEM is clearly problematic for inverse problems in which the data is informative since then the bulk of the prior distribution may have little overlap with the bulk of the posterior distribution, so prior samples used to tune the EEM could carry virtually no information about the posterior distribution\footnote{It follows from the optimality criteria noted in Section~\ref{sec:view} that, in the limit of infinitely many tuning samples, the prior-tuned EEM is a best approximation to the prior distribution.}. Using this \textit{a priori} EEM in DA leads to only modest improvements in computational efficiency, and typically is not significantly more efficient than unmodified MH; see Section~\ref{sec:1d}.  Instead, Cui \emph{et al.} developed the adaptive delayed acceptance (ADA) Metropolis--Hastings (MH) MCMC algorithm~\cite{wrr_cfo_2011}, that tunes the EEM by estimating \emph{posterior} statistics of $\B(\bx)=F(\bx)-F^*(\bx)$. This appears to present a ``chicken-and-egg" problem~\cite{wiki:cote}, in which it is necessary to have explored the posterior distribution to tune the randomizing distribution in order to explore the posterior distribution. They resolved this dilemma by \emph{adapting} the parameters in the EEM, while running the MCMC sampler. In doing so, they followed the advice given by Jeff Rosenthal that: ``if there is some property of the target distribution that you want but don't have, then adapt to it''~\cite{RosenthalParkCity}. A practical algorithm implementing this method is presented in Section~\ref{sec:ada}. A proof of ergodicity was presented in~\cite{CuiFoxO'Sullivan2019}; we present a more succinct proof of ergodicity in Appendix~\ref{sec:ergo} that should appeal to mathematicians.  As noted in~\cite{wrr_cfo_2011}, the resulting adaptive algorithm is significantly faster than non-adaptive DA using the \textit{a priori} EEM in all computational measures, since the off-line calculation is avoided and the resulting MCMC is more statistically efficient. 

A second 
computational efficiency made possible by using DA, and also ADA, is the use of \emph{state-dependent} approximations $F_{\bx}^*(\cdot)$. 
As we demonstrate in Section~\ref{sec:compex}, and has been demonstrated in other contexts~\cite{quiroz2018}, the use of a state-dependent approximation is necessary for best improvement in computational efficiency. In particular, the state-dependent approximation $F_{\bx}^*(\cdot)$ that results from a local zeroth-order correction of a fixed approximation $F^*(\cdot)$ has zero mean in the EEM, i.e., $\boldsymbol\mu_{\bB}=\boldsymbol{0}$; see Section~\ref{sec:approx}. As noted in~\cite{KaipioSomersalo2004}, accurately estimating the mean of the EEM has greater effect on the quality of approximations than does accurately  estimating the variance, implying that this state-dependent approximation without the EEM already achieves better posterior approximation than does a state-independent approximation with a not-optimally-tuned EEM.

The combination of adapting the randomizing distribution and performing the local zeroth-order correction means that both the reduced model and randomizing distribution depend on the iteration number of the MCMC in ADA. These are denoted $F^*_{\bx_n}$ and $\theta_{\bB,n}$, for iteration $n$ in the DAG in Fig.~\ref{fig:dagipBx} that attempts to display the resulting probabilistic model.
\begin{figure}[h]
   \centerline{\includegraphics[scale=1]{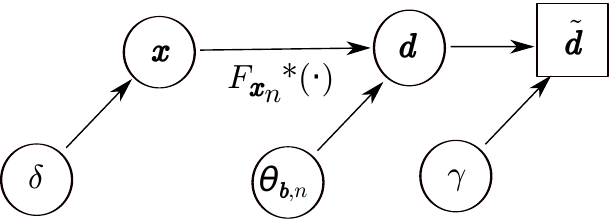}}
   \caption{A DAG representing the conditional structure in ADA, in which both the reduced model $F_{\bx}^*$ and randomization parameters $\theta_{\bB}$ depend on the state at iteration number $n$ of the MCMC.\label{fig:dagipBx}}
\end{figure}

\subsubsection{A Change of Viewpoint}
\label{sec:view}

All the methods we have described so far build a stochastic model to the error introduced by an approximate forward map, based on modeling principles, 
and then use that model as the random correction to the approximate forward map in Eqn~\eqref{eq:addrand}. This is the titular randomizing. However, we wish to present an alternative, mathematically equivalent, interpretation of the resulting algorithms that motivates further computational efficiencies. 

For simplicity of exposition, consider the simple case of Gaussian distributions that are independent in components so that $\mathbf{\Sigma}_{\be}=\sigma^2 I$ and $\mathbf{\Sigma}_{\bB}=\lambda^2 I$,  giving the approximated likelihood function in Eq~\eqref{eq:bayesstar}
\begin{equation}
 L^*(\tilde{\bd} | {\bx}, \sigma,\theta_{\bB})   \propto \exp\left[ -\frac{1}{2}  \frac{\| F^*({\bx})+\boldsymbol\mu_{\bB}-{\tilde{\bd}} \|^2}{2(\sigma^2 + \lambda^2)}  \right],
 \label{eq:likestar}
\end{equation}
where $\theta_{\bB}=(\mu_{\bB},\lambda^2)$.
The reason that the EEM is computationally feasible, as with the error models in~\cite{KennedyO'Hagan2000,KennedyO'Hagan2001,higdon2006}, is that this modified likelihood function~\eqref{eq:likestar} is not significantly more expensive to calculate than the original functional form (that has $\boldsymbol\mu_{\bB}=\boldsymbol{0}$ and $ \lambda^2=0$), whatever the values of $\boldsymbol\mu_{\bB}$ and $ \lambda^2$. Indeed, the \emph{only} stochastic models for model error that can be considered are those that lead to a computationally feasible modified likelihood function. It therefore seems much more straightforward to view such modifications directly, rather than starting with stochastic error models constrained to those models that give computable modified likelihood functions. Accordingly, we forget about the random models for errors and simply view the formula in~\eqref{eq:likestar} as a family of modified likelihood functions, parametrized by $\boldsymbol\mu_{\bB}$ and $ \lambda^2$, and ask: \emph{how should we choose the modification parameters $\boldsymbol\mu_{\bB}$ and $ \lambda^2$ to best approximate the true posterior distribution}? That is, we choose the free parameters in~\eqref{eq:likestar} according to some criterion of optimality that fits the approximated posterior to the true posterior distribution.

We do not propose a single criterion of optimality here as useful criteria will likely depend on the modifications used, while  determining actual computational efficiency needs to be performed within the MCMC, as discussed later and in Section~\ref{sec:betabar}. However, there are some obvious contenders for tuning parameters in the present setting: One such is minimizing the Kullback--Leibler divergence of the approximating distribution to the true posterior distribution, $\operatorname{KL}(\pi_\text{post}\| \pi^*)$. Since both the true posterior $\pi_\text{post}$ in Eqn~\eqref{eq:post} and the approximate target $\pi^*$ in Eqn~\eqref{eq:bayesstar} with approximate likelihood function~\eqref{eq:likestar} (see also Eqn~\eqref{eq:a_enhan_post}) are in exponential form with common base measure equal to the prior distribution $\pi_\text{prior}$, we may draw on the standard result that $\operatorname{KL}(\pi_\text{post}\| \pi^*)$ is minimized when the expected sufficient statistics of each distribution are equal, often called \emph{moment matching}, see, e.g., ~\cite[Section 10.7]{bishop2006}. If we compute expectations over $\pi^*$ using the DA MCMC chain then the result is that we must choose $\boldsymbol\mu_{\bB}$ and $\lambda^2$ to be precisely the posterior statistics of the approximation error $\B(\bx)=F(\bx)-F^*(\bx)$ that we use in ADA; see Section~\ref{sec:ppaem}. The same choice of parameters also results from a second plausible criterion, that is to choose the parameters so that the expected logarithm of the acceptance probability in Step $2$ of  DA is close to 0, which corresponds to the approximation being good on average; see Section~\ref{sec:da}. In Section~\ref{sec:ect} we use a least-squares optimality criterion for one parameter. 

This provides a principle for more general corrections: The functional form of the likelihood function is modified, by introducing parameters in any computationally-cheap way that seems reasonable, for evaluating the approximate likelihood with the approximate forward map, $F^*$, and parameters are set to values that optimize a suitable criterion of fit to the true posterior distribution. As we show in Section~\ref{sec:betabar}, the quality of the resulting approximation may be quantitatively evaluated by calculating the acceptance rate in Step~\ref{step:da2} of DA, thereby validating such modifications, whatever the motivation for the modification and criterion of optimality. 

For example, one such modification is to also multiply the output of the reduced model $F^*(\bx)$ by a diagonal matrix, to give the modified approximate likelihood
\begin{equation}
 L^*(\tilde{\bd} | {\bx}, \sigma, \theta_{\bB})   \propto \exp\left[ -\frac{1}{2}  \frac{\| \rho F^*({\bx})+\boldsymbol\mu_{\bB}-{\tilde{\bd}} \|^2}{2(\sigma^2 + \lambda^2)}  \right]
 \label{eq:likestar1}
\end{equation}
where the multiplier $\rho$ mimics the regression coefficient used in~\cite{KennedyO'Hagan2000}. In the ECT example in Section~\ref{sec:ect} this factor is introduced to compensate for the error in Neumann boundary conditions when using a coarse FEM discretization. The modified likelihood is used, with optimized parameters $\theta_{\bB}=\{\rho,\boldsymbol\mu_{\bB},\lambda^2\}$, to perform posterior inference in ECT in a circular region with $F^*$ defined by a FEM discretization using only $76$ elements; this is substantially fewer than the $900$ elements used as the approximate model with \textit{a priori} EEM in a related EIT problem in~\cite{KaipioSomersalo2004}, so results in a huge reduction in computational cost for a similar quality of approximation.

Numerical analysts may feel more comfortable with this new interpretation, compared to the idea of randomizing the reduced model using `Bayesian' modelling principles, since now the approximate reduced model is used with a (deterministic) modified likelihood calculation, and the modification parameters are chosen by a (deterministic) optimality criterion, that may be validated by online estimates of relative efficiency. 
However, as we have noted, the two interpretations are mathematically equivalent. 

While we have implied that \emph{any} computationally-cheap modification to the likelihood function is allowable, a technical restriction when adapting parameters using ADA is that adapting to the optimal parameter value must be possible in a way that achieves diminishing adaptation, to ensure the conditions for ergodicity are satisfied; see Theorem~\ref{theo3}. This is straightforward for the mean ${\boldsymbol \mu}_{\bB}$ and covariance $\mathbf{\Sigma}_{\bB}$ of the EEM, using standard estimators for these quantities; see Section~\ref{sec:approx}. This issue needs to be resolved for each type of modification. It also seems desirable (to us) that the original functional form of the likelihood be recovered for some value of parameters, e.g., $\{\rho,\boldsymbol\mu_{\bB},\lambda^2\}=\{1,\boldsymbol{0},0\}$ in~\eqref{eq:likestar1}. 

}

\subsection{Contents of this Paper}

\cf{As indicated in Section~\ref{sec:view}, this paper introduces two innovations compared to our previous work in this area~\cite{wrr_cfo_2011,CuiFoxO'Sullivan2019}. The first is the new principle for correcting the error introduced by an approximate forward map, away from the idea of stochastic modelling of the error and forming the likelihood function for a randomized reduced model, to the simpler and more general principle of directly modifying the likelihood function with tuning of parameters to best approximate the posterior distribution. This opens up new possibilities for `correcting' the approximation to the forward map and, hence, leads to better approximations and/or cheaper computation for a given level of approximation. The second is leveraging the result established in~\cite{FoxEnumath2017} that the acceptance rate in Step~\ref{step:da2} of DA (Alg.~\ref{alg:da}) quantitatively accesses the quality of an approximation; see Section~\ref{sec:da}. For a given reduced forward map, the quality of the approximation achieved by a particular modification to the likelihood function is evaluated by computing this statistic over the DA chain.

Under appropriate assumptions these ideas lead to the same calculations as given by previous application of the EEM, and so the examples presented in~\cite{wrr_cfo_2011,CuiFoxO'Sullivan2019} are relevant; accordingly we briefly present those examples in Section~\ref{sec:geo}, highlighting the posterior statistics that we now understand provide quantitative assessment of the various modified likelihood functions.
A more general correction is utilized in the new example that we present in Section~\ref{sec:ect} in which we explore the limits of how coarse a FEM discretization may be while still providing useful estimates in the inverse problem of ECT; the quantitative assessment shows that we may use significantly fewer elements than has been previously demonstrated using EEM.
}

The remainder of this paper is structured as follows:
Section \ref{sec:pe} reviews the basics of sample-based inference and existing algorithms including delayed-acceptance and adaptive algorithms. 
Section \ref{sec:approx} presents deterministic and randomized approximations to the forward map, the induced approximate likelihood functions and posterior distributions, and the ADA algorithm that utilizes these approximations. 
Section~\ref{sec:compex} presents computational studies. Section~\ref{sec:geo} presents two case studies of using ADA to calibrate geothermal reservoir models, taken from~\cite{wrr_cfo_2011,CuiFoxO'Sullivan2019}; here we give very brief details of the inverse problems, and highlight the computed results that pertain to the quantitative measure of efficiency in Section~\ref{sec:betabar}. The first is a 1D homogeneous model with 7 unknown parameters, and uses synthetic transient data. The second example predicts the hot plume of a 3D multi-phase geothermal reservoir model  with $10,049$ unknown parameters by estimating the heterogeneous and anisotropic permeability distribution and the heterogeneous boundary conditions. 
Section~\ref{sec:ect} presents a new case study in ECT of how coarse an approximation can be while still being usable. The intent is to generate and tune an extremely coarse discretization that could be used standalone for online inference.
Section \ref{sec:dis} summarizes and discusses results in the paper, including a comparative discussion of~\cite{calvetti2018iterative}. Appendix~\ref{sec:ergo} presents a new compact proof of ergodicity of ADA when using the approximations developed in Section~\ref{sec:approx}.

\section{Posterior Exploration}
\label{sec:pe}

In this section we review existing sample-based Bayesian methods and algorithms for inverse problems that are relevant to the algorithms and randomized reduced models developed in Section~\ref{sec:approx}.

\subsection{Sample-based Inference}

Sample-based inference proceeds by computing Monte Carlo estimates of posterior statistics, to give `solutions' and quantified uncertainties, using samples drawn from the posterior distribution via Markov chain Monte Carlo (MCMC) sampling.

Estimates of parameters and model predictions can be calculated as Monte Carlo estimates of the expected value of those quantities over the posterior distribution. For quantity $g({\bx})$, the estimate, denoted $\overline{g}_N$, is defined by 
\begin{equation}
\E[g] = \int g({\bx}) \pi({\bx | {\tilde{\bd}}}) \,\dd{\bx}
                     \approx \overline{g}_N = \frac{1}{N}\sum_{i=1}^{N} g({\bx}_{i}),
\label{eq:expect}
\end{equation}
using $N$ samples drawn from the posterior distribution, i.e., ${\bx}_{i}\sim \pi_\text{post}({\bx} | {\tilde{\bd}})$. 
In this way, the task of estimating parameters or predictive values is reduced to the task of drawing samples from the posterior distribution; this defines sample-based Bayesian inference. We present algorithms for sampling from $\pi_\text{post}$ in Sections~\ref{sec:mh} and \ref{sec:da}, and a novel efficient algorithm in Section~\ref{sec:ada}.

\subsection{Metropolis--Hastings Dynamics}
\label{sec:mh}
All sampling methods we develop in this paper are based on the Metropolis--Hastings (MH) algorithm~\cite{metropolis,Hastings1970}.
This algorithm simulates a Markov chain of random variables, that converge in distribution to the posterior distribution $\pi_\text{post}({\bx}|\tilde{\bd})$ as the number of iterations $n\rightarrow\infty$. One initializes the chain at some starting state ${\bx}_0$, usually drawn from an over-dispersed distribution, 
then iterate as in Alg.~\ref{alg:mh}.
\begin{algorithm}
\caption{Metropolis--Hastings~\cite{metropolis,Hastings1970} targeting $\pi_\text{post}({\bx}|{\tilde{\bd}})$}
At iteration $n$, given ${\bx}_n = {\bx}$, ${\bx}_{n+1}$ is determined by:
\begin{enumerate}
\item Propose new state ${\by}$ from some distribution $q\left(\cdot|{\bx}\right)$.
\item With probability \vspace*{-1em}
\[
      \min\left\{1, \frac{\pi_\text{post}({\by}|{\tilde{\bd}}) q(\bx|\by)}
                 {\pi_\text{post}({\bx}|{\tilde{\bd}})q(\by|\bx)}   
      \right\}
\]
set ${\bx}_{n+1} = {\by}$, otherwise ${\bx}_{n+1} = {\bx}$.
\end{enumerate}
\label{alg:mh}
\end{algorithm} 
After a burn-in period, in which the chain effectively loses dependency on the starting state, the MH algorithm produces a sequence of correlated samples distributed as $\pi_\text{post}({\bx}|\tilde{\bd})$.

Samples from the chain, typically after burn-in is discarded, may be substituted directly into the Monte Carlo estimate in Eqn~\eqref{eq:expect} to produce the estimate $\overline{g}_N$ of quantity $\E[g]$. The rate at which $\overline{g}_N\stackrel{\cal D}{\rightarrow}\E[g]$ depends on the degree of correlation~\cite{Sokal,mcmc_geyer1992}; chains that are fast to converge have lower correlation between adjacent samples. Total compute time equals the compute cost per iteration multiplied by the number of iterations required to achieve the desired tolerance; different MCMC algorithms will differ in both these measures, so it is necessary to consider both these measures, as detailed in Sections~\ref{sec:se} and \ref{sec:ce}.

The only choice one has in Alg.~\ref{alg:mh} is the choice of proposal distribution $q(\cdot|\cdot)$; the choice is largely arbitrary, though has a significant influence on the rate of  convergence. Traditionally, the proposal distribution is chosen from some simple family of distributions, then manually tuned in a ``trial and error'' manner to optimize the rate of convergence. We present automatic, adaptive methods for tuning the proposal in Sections~\ref{sec:adapt} and \ref{sec:ada} (in addition to the adapted randomized reduced forward models in Section~\ref{sec:approx}).

\subsection{Statistical Efficiency}
\label{sec:se}
Convergence of the Monte Carlo estimate $\overline{g}_N$ to $\E[g]$ is guaranteed by a central limit theorem~\cite{clt_kipnis1986}  that gives
\(\overline{g}_N-\E\left[ g \right]\sim\N\left(0,\Var(\overline{g}_N)\right) \) as $N\rightarrow\infty$. 
When the ${\bx}_{i}$ are independent
\[ \Var(\overline{g}_N) = \frac{\Var(g)}{N}. \]
When the ${\bx}_{i}$ are correlated  (for  large $N$)
\[ \Var(\overline{g}_N) = \frac{\Var(g)}{N}\left(1+2\sum_{i=1}^\infty \rho_{gg}(i) \right)=\tau \frac{\Var(g)}{N}\]
where $\rho_{gg}$ is the autocorrelation coefficient for the chain in $g$. Hence, the rate of variance reduction, compared to independent samples, is reduced by the factor $\tau$
which is called the integrated autocorrelation time (IACT) for the statistic $g$~\cite{Sokal}. 
We can think of $\tau\geq1$ as the length of the correlated chain that produces the same variance reduction as one independent sample. 
We call $1/\tau\leq 1 $ the \emph{statistical efficiency} (higher is better), while $N / \tau$ is the effective (independent) sample size (ESS). 

\subsection{Delayed Acceptance}
\label{sec:da}

Applying standard MH can be computationally costly as each iteration requires evaluating the posterior density, which involves simulating the forward map $F(\by)$ at proposed parameters $\by$, and typically many iterations are required for convergence of estimates. 

We develop sampling algorithms with reduced computational cost by using the framework of the delayed acceptance (DA) MH algorithm of Christen and Fox~\cite{ChristenFox2005} that uses two accept-reject steps; see Alg.~\ref{alg:da}. The first is evaluated using an approximation to the target distribution that can be relatively arbitrary, while the second accept-reject step ensures that the Markov chain correctly targets the desired distribution.
The computational cost per iteration is reduced because only those proposals that are accepted using the approximation  $\pi_{\bx}^*(\cdot)$ go on to evaluation of the posterior distribution $\pi_\text{post}(\cdot)$, that requires evaluating the full, expensive forward map (if $\by$ is rejected at the first step then the chain does not move and no further calculation is required). 

\begin{algorithm}
\caption{Delayed acceptance~\cite{ChristenFox2005} targeting $\pi_\text{post}({\bx}|{\tilde{\bd}})$\label{alg:da}}
Given ${\bx}_n = {\bx}$ and approximate target distribution $\pi_{\bx}^*(\cdot)$, ${\bx}_{n+1}$ is determined by:
\begin{enumerate}
\item \label{step:da1} Propose ${\by}$ from some distribution $q\left(\cdot|{\bx}\right)$. With probability 
\begin{equation*}
  \alpha\left({\bx},{\by}\right) = \min\left\{1, \frac{\pi_{\bx}^*({\by}) q(\bx|\by)}
                     {\pi_{\bx}^*({\bx})q(\by|\bx)}   ,
      \right\}
\end{equation*}
promote ${\by}$ to be used in Step 2, otherwise set ${\by}={\bx}$ 
\item \label{step:da2} The effective proposal distribution at this step is $q^*(\by|\bx)=q(\by|\bx)\alpha({\bx},{\by})$ for $\by\neq\bx$.\\
With probability 
\[
  \beta\left({\bx},{\by}\right) =\min\left\{1, \frac{\pi_\text{post}({\by}|{\tilde{\bd}}) q^*(\bx|\by)}
                     {\pi_\text{post}({\bx}|{\tilde{\bd}})q^*(\by|\bx)}   
      \right\}
\]
set ${\bx}_{n+1} = {\by}$, otherwise ${\bx}_{n+1} = {\bx}$. 
\end{enumerate}
\end{algorithm} 

The DA algorithm allows the approximation to depend on the current state of the MCMC, shown by the notation $\pi_{\bx}^*(\cdot)$, and is guaranteed to converge to the target posterior distribution $\pi_\text{post}({\bx}|{\tilde{\bd}})$ under mild conditions~\cite{ChristenFox2005}. A special case of DA is the surrogate transition method~\cite{mcmc_liu} that requires a fixed, `surrogate' approximate target distribution, i.e, $\pi^*({\by})$ does not depend on the current state $\bx$. Using an approximation that depends on the state turns out to be necessary for improving computational efficiency in the applications that we consider. 

\cf{
\subsection{Quality of the Approximation is Measured by $\bar{\beta}$}
\label{sec:betabar}

We follow a result in~\cite{FoxEnumath2017}, that builds on the coupling/separation analysis in~\cite{nicholls2012coupled}, to show that the second accept/reject Step~\ref{step:da2} in DA evaluates the quality of the approximation. 

It was shown in~\cite{ChristenFox2005}  that, under mild requirements, $\beta\approx 1$ when $\pi_{\bx}^*(\cdot)\approx \pi(\cdot)$, i.e. the acceptance probability in Step~\ref{step:da2} of DA approaches $1$ when the approximation is good. While this is a cause for optimism, it is not the reverse implication that we need, i.e. that $\beta\approx 1$ implies the approximation is good.

Intuitively, we can see that when the acceptance rate in the second step is very close to $1$, i.e. $\bar{\beta}\approxeq 1$, the second accept/reject step is almost redundant and it might be possible to solely use the approximation for performing inference, i.e. run the MH Alg.~\ref{alg:mh} using $\pi_{\bx}^*(\cdot)$. However, it is not clear that this chain even has an equilibrium distribution when the approximation is state dependent.

The reverse implication, that $\bar{\beta}\approx 1$  does imply that the approximation is good, was proved in~\cite{FoxEnumath2017}. More formally,~\cite{FoxEnumath2017} compared Monte Carlo estimates of the posterior expected value of some statistic $g(\bx)$, denoted $\overline{g}_N$ when evaluated over $N$ steps of a convergent DA chain targeting $\pi$, and denoted $\overline{g}^*_N$ when evaluated over the chain that omits the accept/reject in Step~\ref{step:da2}, to give the following theorem:
\begin{theo}[\cite{FoxEnumath2017}]
If the proposal is such that the expected square jump size is uniformly bounded, i.e.,
\(
  \E_q[\|\bx'-\bx\|^2] \leq M<\infty,
\)
and $\|g\|$ is bounded by a uniformly continuous function, then $E[\|\bar{g}^*_N-\bar{g}_N\|] {\rightarrow} 0$, i.e., $\bar{g}^*_N \rightarrow \bar{g}_N$ in expectation, as $\E[\beta]\rightarrow 1$, and with the same rate.
\end{theo}
That is, any estimate computed over the chain of length $N$ using the approximate posterior will converge to the estimate computed using a chain of length $N$ using the exact posterior distribution (and that estimate converges to the true value due to ergodicity of the chain targeting the correct posterior distribution as $N\rightarrow\infty$) as the acceptance rate in Step~\ref{step:da2} of DA approaches $1$. This is the sense in which we say an approximation is `good'.
This is a practically useful result as an estimate of the acceptance rate in Step~\ref{step:da2} of DA, $\bar{\beta}$, may be evaluated over the chain to determine the quality of the approximation $\pi_{\bx}^*(\cdot)$.

This result motivates our second computed example in Section~\ref{sec:ect} in which a state-independent reduced model is used with a modified likelihood that is tuned to increase $\bar{\beta}$ with the aim of using the cheap approximation for sample-based inference in place of the expensive true posterior distribution.
}

\subsection{Computational Efficiency}
\label{sec:ce}
We define the {\it computational efficiency} of a sampler to be the ESS per CPU time. Hence, from Section~\ref{sec:se}, this is proportional to the variance reduction in estimates per CPU time.

The DA Alg.~\ref{alg:da} necessarily has lower statistical efficiency than the unmodified counterpart in Alg.~\ref{alg:mh}~\cite{ChristenFox2005}. That is, $\tau_\text{DA}\geq\tau_\text{MH}$ for any quantity $g$, so more steps of DA are required than of MH to evaluate estimates to a desired accuracy.
Fortunately, DA may still be more computationally efficient than the standard MH. 
Let $t^*$ and $t$ be the CPU time to evaluate the approximate and exact posterior density, respectively, and let the 
average acceptance probability in Step~\ref{step:da1} of DA  be denoted $\bar{\alpha}$.
Then, the increase in computational efficiency of DA compared to standard MH is the ratio of ESS for fixed CPU time~\cite{CuiFoxO'Sullivan2019}
\begin{equation}
 \frac{\mbox{ESS}_{\mathrm{DA}}}{\mbox{ESS}_{\mathrm{MH}}} = 
 \frac{\tau_{\mathrm{MH}}}{\tau_{\mathrm{DA}}}\frac{1}{\bar{\alpha} + t^* / t}.
 \label{eq:su_factor}
\end{equation}
(Of course, ${\mbox{ESS}_{\mathrm{DA}}}/{\mbox{ESS}_{\mathrm{MH}}}<1$ means that DA is less efficient than MH.)
The ratio $\tau_{\mathrm{MH}}/\tau_{\mathrm{DA}} \leq 1$ is the decrease in statistical efficiency, while
$\bar{\alpha} + t^* / t$ gives the decrease in average compute cost per iteration. 
It is necessary to address both factors if computational efficiency is to be increased.
The ideal is to have $\tau_{\mathrm{MH}}/\tau_{\mathrm{DA}}\approx 1$, i.e., statistical efficiency is not decreased, and $t^* / t\approx 0$ which occurs when the approximation is very cheap to calculate.

It can be challenging to balance the reduction in CPU time against accuracy of the reduced model. 
Using a lower accuracy reduced model, which runs faster compared to a more accurate one, will reduce the average CPU time per iteration  of the MCMC, but at the risk of lower statistical efficiency that increases the number of MCMC iterations required.
The framework of DA affords two routes to improving a reduced model, by using the pairing of reduced and exact evaluations at the second step of DA (following first-step acceptances). The first is calculating a local, zeroth-order correction to the reduced model; this is Approximation~\ref{appr:local} in Section~\ref{sec:approx}. The second is to adapt a randomization of the reduced model by adapting to posterior statistics of the reduced-model error;  this is Approximation~\ref{appr:aemsd} in Section~\ref{sec:approx}. Each of these routes increases the acceptance rate in Step~\ref{step:da2} without changing the cost of the reduced model, and hence improves computational efficiency.  Our adaptive algorithms draw on adaptive MCMC, that we review next.

\subsection{Adaptive MCMC}
\label{sec:adapt}
A general class of adaptive algorithms was established by~\cite{RobertsRosenthal2007} with simplified regularity conditions required for ergodicity, namely simultaneous uniform ergodicity and diminishing adaptation.
The ergodicity of many practical adaptive MCMC algorithms can be established using these simplified conditions. 

Provably ergodic adaptive MCMC was initiated by the adaptive Metropolis (AM) algorithm of Haario et al.~\cite{adaptive_haario2001} that adapts the random-walk proposal distribution in a Metropolis algorithm. Almost all subsequent adaptive MCMC algorithms follow this precedent of adapting the proposal distribution, only, as in Alg.~\ref{alg:adapt}.
\begin{algorithm}
\caption{Adaptive proposal Metropolis\label{alg:adapt}}
Given ${\bx}_n = {\bx}$ and symmetric proposal $q_n(\cdot|\bx)$, ${\bx}_{n+1}$ and $q_{n+1}(\cdot|\cdot)$ are determined by:
\begin{enumerate}
\item 
Propose ${\by}$ by drawing $\by\sim q_n(\by|\bx)$
\item With probability $\min\big\{1, \pi_\text{post}({\by}|{\tilde{\bd}}) / \pi_\text{post}({\bx}|{\tilde{\bd}}) \big\}$, ${\bx}_{n+1} = {\by}$, otherwise ${\bx}_{n+1} = {\bx}$.
\item Update proposal $q_{n+1}$.
\end{enumerate}
\end{algorithm}

Specifically, for small $\gamma>0$ AM uses the random-walk Gaussian proposal
\begin{equation}
q_n(\by|\bx) = \left\{ \begin{array}{ll} \N(\by;{\bx},\frac{0.1^2}{d} {\bf I}_d) & n \leq 2d \\ 
         \N(\by;{\bx},(1-\gamma)\frac{2.38^2}{d}\boldsymbol\Sigma_n+\gamma \frac{0.1^2}{d}{\bf I}_d) & n > 2d \end{array} \right.,
\label{eq:am}
\end{equation}
utilizing the empirical covariance $\mathbf{\Sigma}_n$ estimated over the Markov chain.

In practice, the proposal~\eqref{eq:am} does not ensure sufficient statistical efficiency in the problems we consider.  
More effective is the grouped components adaptive Metropolis (GCAM) proposal that uses an AM-type proposal separately for groups of components of $\bx$, with empirical covariance matrix and scale variables, i.e., the coefficients in~\eqref{eq:am},  estimated separately for each group; see ~\cite{CuiFoxO'Sullivan2019} for details.

We extend existing adaptive MCMC algorithms by also adapting the approximate likelihood function, and hence adapt the approximation to the posterior distribution. Those adaptive approximations are presented next in Section~\ref{sec:approx}. We establish ergodicity in Section~\ref{sec:ergo} by following~\cite{RobertsRosenthal2007}.

\section{Approximations to $F(\cdot)$ and $\pi_\text{post}$}
\label{sec:approx}
We assume that we have a deterministic reduced model $F^*(\cdot)$ that approximates $F(\cdot)$, built using one of the methods outlined in Section~\ref{sec:intro}. The notation $F^*(\cdot)$ implies that the reduced model does not depend on the state of the MCMC, as would be the case with a fixed coarse-grid discretization. State-dependent approximations, such as a local linearization, may also be accommodated; see Approximation~\ref{appr:local}, later.

We start with the common approximation to the posterior distribution that simply uses $F^*(\cdot)$ in place of the true forward map $F(\cdot)$.
\begin{appr}
\label{appr:coarse}
Approximate posterior distribution using $F^*(\cdot)$ in place of $F(\cdot)$:
\begin{equation}
\pi^*({\bx}|\tilde{\bd}) \propto \exp\left[ -\frac{1}{2}  \{ F^*(\bx)-{\tilde{\bd}} \}^T \mathbf{\Sigma}_{\be}^{-1} \{ F^*({\bx})-{\tilde{\bd}} \} \right] \pi_\mathrm{prior}(\bx).
\label{eq:coarse_post}
\end{equation}
\end{appr}
The approximation in~\eqref{eq:coarse_post}, by itself, can result in biased estimates while producing uncertainty intervals that are too small~\cite{KaipioSomersalo2004,KaipioSomersalo2007}. This indicates that the approximate posterior has displaced support and is too narrow to include the support of the accurate posterior.

Given a reduced model $F^*(\cdot)$, Eqn~\eqref{eq:cali} can be rewritten
\begin{eqnarray}
{\tilde{\bd}} & = & F^*({\bx}) + \{ F({\bx}) - F^*(\bx) \} + \be\nonumber \\
& =  & F^*({\bx}) + \B({\bx}) + \be.
\label{eq:enhanm}
\end{eqnarray}
By assuming that the model reduction error $\B$ can be modelled as independent of the model parameters and is Gaussian, \cite{KaipioSomersalo2007} introduced the enhanced error model (EEM) (cf. Eqn~\eqref{eq:addrand})
\[
{\tilde{\bd}}  =  F^*({\bx}) + \bB + \be,
\]
where $\bB \sim N({\boldsymbol \mu}_{\bB}, \mathbf{\Sigma}_{\bB})$. Improved point estimates in~\cite{KaipioSomersalo2007} indicate that the EEM plausibly improves the approximation of the posterior distribution compared to Approximation~\ref{appr:coarse}. 
\begin{appr}
\label{appr:aemsi}
Approximate posterior distribution using the reduced model $F^*(\cdot)$ and EEM:
\begin{equation}
\pi^*({\bx}|\tilde{\bd}) \propto \exp\left[ -\frac{1}{2}  \{ F^*({\bx})+\boldsymbol\mu_{\bB}-{\tilde{\bd}} \}^T (\mathbf{\Sigma}_{\bB}+\mathbf{\Sigma}_{\be})^{-1} \{ F^*({\bx})+\boldsymbol\mu_{\bB}-{\tilde{\bd}} \} \right] \pi_\mathrm{prior}({\bx}).
\label{eq:a_enhan_post}
\end{equation}
\end{appr}

\subsection{Prior and Posterior Error Models}
\label{sec:ppaem}
The EEM was estimated {\it a priori} in \cite{KaipioSomersalo2007}, before utilizing data and solving the inverse problem, resulting in the estimates for mean and covariance of $\bB$
\begin{eqnarray}
\boldsymbol\mu_{\bB} & = & \int_{\mathcal{X}} \B({\bx}) \pi_\text{prior}({\bx})\, \dd{\bx} 
       \approx  \frac{1}{L}\sum_{i = 1}^{L} \B({\bx}_i),  \label{eq:muprior}\\
\mathbf{\Sigma}_{\bB} & = & \int_{\mathcal{X}} \{ \B({\bx}) - \boldsymbol\mu_{\bB} \} \{ \B({\bx}) - \boldsymbol\mu_{\bB} \}^T \pi_\text{prior}({\bx})\, \dd{\bx} \nonumber\\
       & \approx & \frac{1}{L-1}\sum_{i = 1}^{L} \{ \B({\bx}_i) - \boldsymbol\mu_{\bB} \}\{ \B({\bx}_i) - \boldsymbol\mu_{\bB} \}^T, \label{eq:sigmaprior}
\end{eqnarray}
where $\B({\bx})=F({\bx}) - F^*(\bx)$ and ${\bx}_i \sim \pi_\text{prior}({\cdot}), i = 1,\cdots,L$, are $L$ samples drawn from the prior distribution.
This \textit{a priori} EEM could be far from optimal over the support of the posterior distribution even though it may fit the prior distribution, requires appreciable pre-computation, and gives only a small improvement in computational efficiency; see Section~\ref{sec:1d}.

We make a better approximation to the posterior distribution by estimating the EEM over the posterior distribution. That is, we evaluate
\begin{eqnarray}
\boldsymbol\mu_{\bB} & = & \int_{\mathcal{X}} \B({\bx}) \pi_\text{post}({\bx} \mid \tilde{\bd}) \,\dd{\bx} , \label{eq:postmu} \\
\mathbf{\Sigma}_{\bB} & = & \int_{\mathcal{X}} \{\B({\bx}) - \boldsymbol\mu_{\bB}\}\{\B({\bx}) - \boldsymbol\mu_{\bB}\}^T \pi_\text{post}({\bx} \mid \tilde{\bd}) \,\dd{\bx}. \label{eq:postsigma} 
\end{eqnarray}
This is achieved by evaluating $\B({\bx})$ within the DA Alg.~\ref{alg:da}, and applying adaptive MCMC methods to ensure that the \textit{a posteriori} estimates of $\boldsymbol\mu_{\bB}$ and $\mathbf{\Sigma}_{\bB}$ converge to the values given in Eqns~\eqref{eq:postmu} and \eqref{eq:postsigma}. 
In all computational experiments we find that building the EEM over the posterior leads to better statistical efficiency in the MCMC, than when the EEM is estimated over the prior, so gives a more computationally efficient MCMC that also does not require any precompution.

\subsection{State-dependent Approximations and Error Models}
\label{sec:statedepend}
The work of \cite{ChristenFox2005} demonstrated DA using a local linearization of the forward map as the approximate forward map, which is a local reduced model that depends on the current state ${\bx}$ of the MCMC. 
When using a state-independent reduced model within DA, it is advantageous to make a zeroth-order local improvement 
by using the values of $F({\bx})$ and $F^*({\bx})$ for points  ${\bx}$ that are accepted, and hence become the state of the chain.
Thus, we define the deterministic state-dependent reduced model, as follows.
\begin{appr}
\label{appr:local}
State-dependent reduced model and approximate posterior distribution: Suppose that at iteration $n$, the Markov chain has state ${\bx}_n = {\bx}$. For a proposed state ${\by} \sim q(\cdot|{\bx})$, the state-dependent reduced model  $F^*_{\bx}(\cdot)$ is 
\begin{equation}
F^*_{\bx}({\by}) = F^*({\by}) + \{F({\bx}) - F^*({\bx})\}.
\label{eq:rom_l}
\end{equation}
The resulting approximate posterior distribution is
\begin{equation}
\pi^*_{{\bx}}({\by}|\tilde{\bd})  \propto \exp\left[-\frac{1}{2}  \{F^*_{\bx}({\by})-{\tilde{\bd}}\}^T \mathbf{\Sigma}_{\be}^{-1}\{F^*_{\bx}({\by})-{\tilde{\bd}}\}\right] \pi_\mathrm{prior}({\by}).
\end{equation}
\end{appr}
The zeroth-order correction~\eqref{eq:rom_l} comes at no extra cost as  $F({\bx})$ has already been evaluated when the state $\bx$ was previously accepted.

Let $\B_{\bx}({\by}) = F({\by}) - F^*_{\bx}({\by})$. The state-dependent reduced model~\eqref{eq:rom_l} has the desirable property that $F^*_{\bx}({\bx})=F({\bx})$ and  $\B_{\bx}({\bx}) = {\bf 0}$, in common with local linearization. 

The error induced by the state-dependent reduced model~\eqref{eq:rom_l} can also be estimated by employing the EEM. In particular,  Approximation~\ref{appr:aemsi} and \ref{appr:local} can be combined, at no significant increase in computational cost, to give a more accurate approximation to the posterior distribution. 
\begin{appr}
\label{appr:aemsd}
EEM built over the posterior distribution with a state-dependent reduced model: Suppose that at iteration $n$ the Markov chain is at state ${\bx}_n = {\bx}$ and a proposed state is ${\by} \sim q(\cdot|\bx)$.  The state-dependent approximate posterior distribution is given by
\begin{equation}
\pi^*_{\bx}({\by}|\tilde{\bd}) \propto  \exp\left[-\frac{1}{2}  \{F^*_{\bx}({\by})+\boldsymbol\mu_{\bB}-{\tilde{\bd}}\}^T (\mathbf{\Sigma}_{\bB}+\mathbf{\Sigma}_{\be})^{-1}\{F^*_{\bx}({\by})+\boldsymbol\mu_{\bB}-{\tilde{\bd}}\}\right] \pi_\mathrm{prior}({\by}).
\label{eq:a_enhan_lp}
\end{equation}
\end{appr}

The mean and covariance of the EEM in Approximation \ref{appr:aemsd} with reduced model \eqref{eq:rom_l} are  
\begin{equation}
  \boldsymbol\mu_{\bB}=\textrm{E}_{\pi({\bx})}\left[\int_{\mathcal{X}} \B_{\bx}({\by}) K({\bx, \by})\, \dd{\by}\right]
  \label{eq:meanAEEM}
\end{equation}
 and 
\begin{equation}
   \mathbf{\Sigma}_{\bB}=\textrm{Cov}_{\pi({\bx})}\left[\int_{\mathcal{X}} \B_{\bx}({\by}) K({\bx, \by})\, \dd{\by}\right],
   \label{eq:covAEEM}
\end{equation}
respectively, where $K({\bx, \by})$ denotes the transition kernel implemented by the MCMC iteration.
The mean of the EEM~\eqref{eq:meanAEEM} for reduced model \eqref{eq:rom_l} can be shown to be ${\bf 0}$, by expanding $\B_{\bx}({\by})=[F({\by}) - F^*({\by})] - [F({\bx}) - F^*({\bx})]$;
\begin{eqnarray*}
  \E_{\pi({\bx})}\left[\int \B_{\bx}({\by}) K({\bx, \by}) \,\dd{\by}\right]  
  & 
  = \int_{\mathcal{X}} \int_{\mathcal{X}} \{F({\by}) - F^*({\by})\} \pi({\bx}) K({\bx, \by}) \,\dd{\by} \,\dd{\bx} \\ 
  & 
  - \int_{\mathcal{X}} \int_{\mathcal{X}} \{F({\bx}) - F^*({\bx})\} \pi({\bx}) K({\bx, \by}) \,\dd{\by} \,\dd{\bx}
\end{eqnarray*}
with the two terms on the right canceling because the kernel $K$ satisfies the detailed balance condition $\pi({\bx}) K({\bx, \by}) =  \pi({\by}) K({\by, \bx})$. 
Accordingly, we set $\boldsymbol\mu_{\bB}={\bf 0}$ in~\eqref{eq:a_enhan_lp}.
The covariance~\eqref{eq:covAEEM} can be computed adaptively at iteration $n$ by the inductive formula
\begin{equation}	
	\mathbf{\Sigma}_{\bB,n} =  \frac{1}{n-1} \left\{ (n-2)\mathbf{\Sigma}_{\bB,n-1}  + \B_{{\bx}_{n-1}}({\bx}_{n})\B_{{\bx}_{n-1}}({\bx}_{n})^T\right\}.
	\label{eq:appr41}
\end{equation}
The approximate posterior distribution \eqref{eq:a_enhan_lp} after $n$ steps of adaptive updating is then
\begin{equation}
\pi^*_{n,{\bx}}({\by}|\tilde{\bd})   \propto  \exp\left[-\frac{1}{2}  \{F^*_{\bx}({\by})-{\tilde{\bd}}\}^T (\mathbf{\Sigma}_{\bB,n}+\mathbf{\Sigma}_{\be})^{-1}\{F^*_{\bx}({\by})-{\tilde{\bd}}\}\right] \pi_\text{prior}(\bx).
\label{eq:a_enhan_ll1}
\end{equation}

These adaptive computations may be evaluated within the adaptive delayed acceptance (ADA) MCMC algorithm, described next.

\subsection{Adaptive Delayed Acceptance Algorithm}
\label{sec:ada}

The ADA MCMC algorithm uses the basic structure of DA but includes adaptivity in both the proposal distribution and in the state-dependent approximate target distribution, allowing each to depend on the current MCMC iteration, as shown in Alg.~\ref{alg:ada}. 
\begin{algorithm}[h]
\caption{Adaptive delayed acceptance Metropolis--Hastings (ADA) }
\label{alg:ada} 
At iteration $n$, given ${\bx}_n = {\bx}$, adapted proposal $q_n(\cdot)$, and approximate target distribution $\pi^*_{n}({\cdot})$, then ${\bx}_{n+1}$ and updated distributions are determined as follows:
\begin{enumerate}
\item Generate a proposal ${\by}\sim q_n(\cdot)$. \label{step:ada1} With probability
\begin{equation*}
\alpha_n({\bx} ,{\by}) = \min\left\{ 1, \frac{\pi^*_{n}({\by})q_n({\bx})}{ \pi^*_{n}({\bx})q_n(\by) }\right\} 
\end{equation*}
 promote ${\by}$ to be used as a proposal for the following step. Otherwise set $\by =\bx$ and proceed. 
\item
The proposal distribution at this step is $q^*_n(\by|\bx) = \alpha_n(\bx,\by)q_n({\by})$ for $\bx\neq\by$. 
With probability
\begin{equation*}
 \min\left\{ 1, \frac{\pi({\by}|\tilde{\bd})q^*_n({\by} ,{\bx})}{\pi({\bx}|\tilde{\bd})q^*_n({\bx},{\by})} \right\} 
\end{equation*} 
set ${\bx}_{n+1}= {\by}$.  Otherwise set ${\bx}_{n+1}=\bx$.
\label{step:ada2}
\item Update the approximation $\pi^*_{n+1}({\cdot})$. \label{step:ada3}
\item Update the adaptive proposal $q_{n+1}(\cdot)$. \label{step:ada4}
\end{enumerate}
\end{algorithm}
In this algorithm, the proposal $q_n( \cdot)$ in step~\ref{step:ada1} and its adaptive update in step~\ref{step:ada4} may have the form of any of the adaptive algorithms, such as the AM in Eqn~\eqref{eq:am} or GCAM.
When Approximation~\ref{appr:aemsd} is used in Step~\ref{step:ada1}, updating of the approximate target distribution in Step~\ref{step:ada3} uses the updating rule for $\mathbf{\Sigma}_{\bB}$ in Eqn~\ref{eq:appr41}.

A proof of ergodicity of ADA for  $\pi(\cdot)$, using any of the Approximations~\ref{appr:coarse} to \ref{appr:aemsd}, is presented in Appendix~\ref{sec:ergo}.

\section{Computed Examples}
\label{sec:compex}
\subsection{Fitting of Geothermal Reservoir Models}
\label{sec:geo}
In this section we apply ADA to two calibration problems for geothermal reservoir models taken from~\cite{CuiFoxO'Sullivan2019}. 
The first is a one dimensional radially symmetric model of the feedzone of a geothermal reservoir with synthetic data. This example is small enough that extensive statistics can be computed to evaluate the efficiency of various approximations; summary results are presented in Table~\ref{table:performance}. The second is sampling a large-scale 3D model with measured data. Computational results are presented to highlight the convergence of approximate estimates guaranteed in Section~\ref{sec:betabar}.

For completeness, we briefly present the formulation of the governing equations and numerical simulator for these inverse problems, though we refer the interested reader to~\cite{geo_book,geo_osullivan_1985} for a complete description of the governing equations of multiphase geothermal reservoirs, and to~\cite{wrr_cfo_2011,CuiFoxO'Sullivan2019} for details in these applications.

\subsection{Data Simulation}
\label{sec:data_simu}
Consider a two phase geothermal reservoir (water and vapour) governed by the general mass balance and energy balance equations
\begin{equation}
\frac{d}{dt} \int_{\Omega} M \,dV = \int_{\partial \Omega} Q \cdot {\bf n} \, d\Gamma + \int_{\Omega} q \, dV ,
\label{eq:conserve}
\end{equation}
for the accumulation term in mass ($M_\textrm{m}$) or energy ($M_\textrm{e}$) per unit volume, 
where $\Omega$ is the control volume and $\partial \Omega$ is its boundary. The term, $q_\textrm{m}$ for mass or $q_\textrm{e}$ for energy, represents sources or sinks in $\Omega$, and associated $Q_\textrm{m}$ or $Q_\textrm{e}$ denotes the flux through $\partial \Omega$. 

A set of nonlinear partial differential equations such as multiphase Darcy's law are used to model accumulation and flux terms. For brevity, we express these terms by the simplified functional relationship between the typical parameters and state of the system
\[
\left( \begin{array}{c} M_\textrm{m} \\ M_\textrm{e} \end{array} \right) = f_{\rm M} (\phi; p, T) , \qquad
\left( \begin{array}{c} Q_\textrm{m} \\ Q_\textrm{e} \end{array} \right) = f_{\rm Q} (k, k_\textrm{rl}, k_\textrm{rv}; p, T) ,
\]
where $\phi$ is porosity, $k$ is a diagonal second order permeability tensor in 3-dimensions, and $k_\text{rl}$ and $k_\text{rv}$ are the relative permeabilities. These are the unknown spatially distributed parameters that we wish to determine. The state of system is represented by the spatially distributed pressure and temperature $(p, T)$ for a single phase system, or pressure and vapour saturation $(p, S_\text{v})$ for a two phase system; these are only partially observable through wells. The subscripts l and v represent the liquid and vapour phases, respectively. 
Relative permeabilities $k_\text{rl}$ and $k_\text{rv}$ are introduced to account for the interference between liquid and vapour phases as they move through the rock matrix in the geothermal reservoir, and we use the van Genuchten-Mualem model for relative permeabilities~\cite{vangenuchten}
\cf{
\begin{align*}
 k_\text{rl} & = \begin{cases} 
                    \sqrt{\frac{1-S_\text{v}-S_\text{rl}}{S_\text{ls}-S_\text{rl}}}
                    \left\{1-\left(1-\left(\frac{1-S_\text{v}-S_\text{rl}}{S_\text{ls}-S_\text{rl}}\right)^{1/m} \right)^m \right\}^2 &\mbox{if } 1-S_\text{v} < S_\text{ls} \\
                    1 & \mbox{if } 1-S_\text{v} \geq S_\text{ls} 
                  \end{cases}  \\
 k_\text{rv} & = 1- k_\text{rl}
\end{align*}
as functions of $S_\textrm{v}$ and three hyperparameters $m$, $S_\textrm{rl}$, and $S_\text{ls}$. }
Spatial discretization of~\eqref{eq:conserve} is based on a finite volume method, implemented in the existing Fortran code TOUGH2~\cite{tough2}. 

\subsection{Well discharge test analysis}
\label{sec:1d}
\begin{figure}[ht]
\centering
\includegraphics[width=0.8\textwidth]{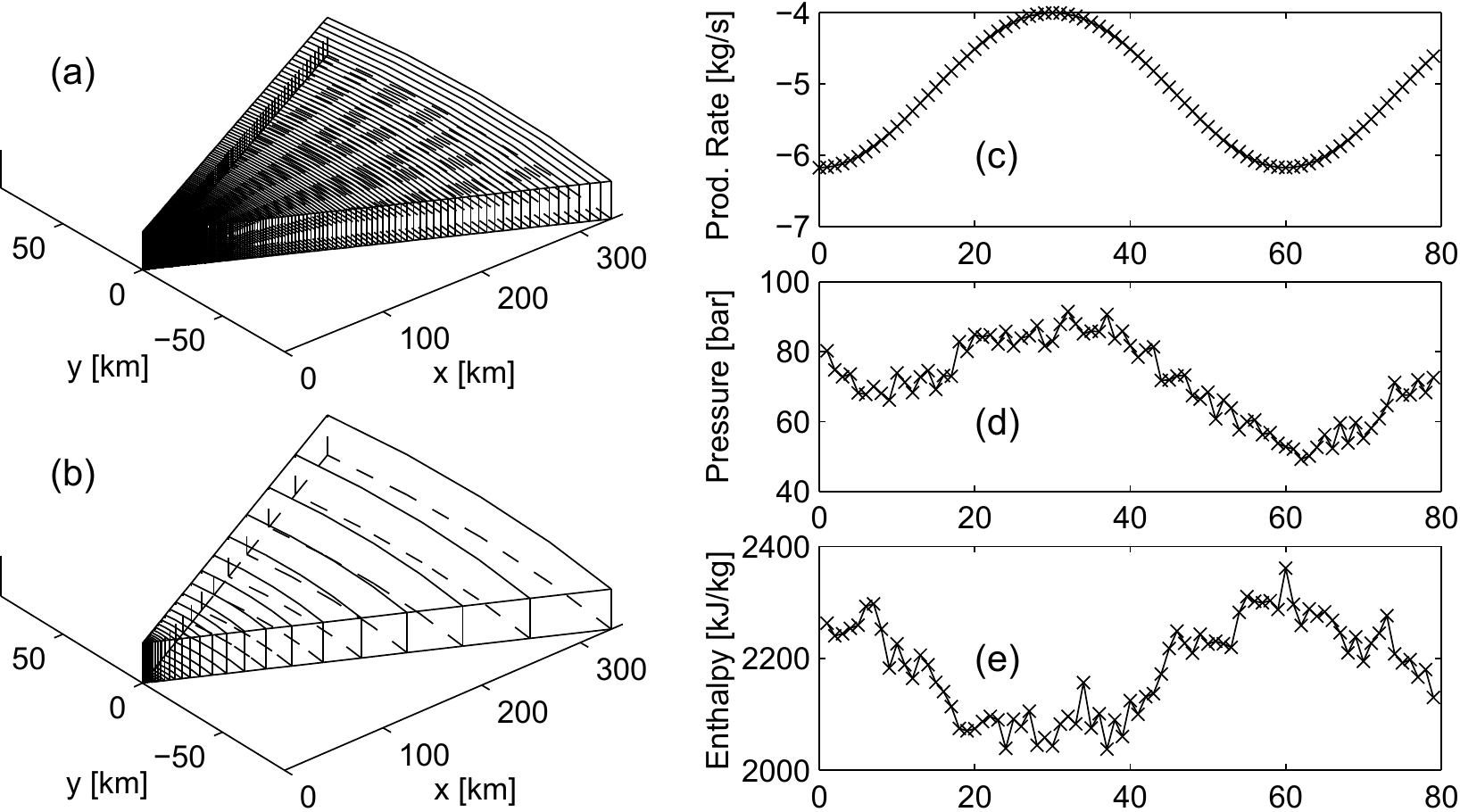}
\caption{Finite volume grids used for well discharge test analysis and data sets used for well discharge test. (a): the forward model (640 blocks), (b): the reduced model (40 blocks), (c): the production rate (kg/second), (d): the pressure (bar), and (e): the flowing enthalpy (kJ/kg).}
\label{figure:wellall} 
\end{figure}

Well discharge analysis is usually used to interpret the near-well properties of the reservoir from pressure and enthalpy data measured during a short period of field production. 
Based on the typical assumption that all flows into the well come through a single layer feedzone, we use a one-dimensional radially-symmetric forward model with 640 blocks as shown in Fig.~\ref{figure:wellall} (a).
A high resolution grid is used immediately outside the well and then cell thickness increases exponentially away from this region. The reduced model uses a coarse grid with 40 blocks, shown in Fig.~\ref{figure:wellall} (b). 
Based on $1,000$ simulations with different sets of parameters on a DELL T3400 workstation, we estimate the CPU time to evaluate the forward model is $2.60$ seconds. CPU time for the reduced model is $0.15$ seconds. 

The seven unknowns required for data simulation are the porosity, permeability (base 10 logarithmic scale), the hyperparameters in the van Genuchten-Mualem relative permeability model, as well as the initial vapour saturation ($S_\textrm{v0}$) and initial pressure ($p_\textrm{0}$) that are used to represent the initial thermodynamic state of the two-phase system:
\begin{equation*}
  {\bx} = \left\{ {\phi}, \log_{10}(k), p_\textrm{0}, S_\textrm{v0},  m, S_\textrm{rl}, S_\textrm{ls}\right\}.
\end{equation*}
These parameters are assumed to be independent and follow non-informative prior distributions with physical bounds~\cite{CuiFoxO'Sullivan2019}. 
The model is simulated over 80 days with production rates varying smoothly from about $4$ kg/second to about $6$ kg/second (see Fig.~\ref{figure:wellall} (c)). Model outputs are pressure ${\bd}_\textrm{p}$ and flowing enthalpy ${\bd}_\textrm{h}$, which defines the forward map
\[
\left(\begin{array}{c} {\bd}_\textrm{h} \\ {\bd}_\textrm{p} \end{array}\right) = F(\bx).
\]
We assume the measurement noise follows i.i.d. zero mean Gaussian distribution with standard deviations $\sigma_p = 3\;\textrm{bar}$ for pressure and $\sigma_h = 30\;\textrm{kJ/kg}$ for the flowing enthalpy. The noise corrupted pressure and flowing enthalpy data are plotted in Figure \ref{figure:wellall} (d) and (e), respectively. 

We ran ADA using Approximation~\ref{appr:coarse}, Approximation~\ref{appr:aemsi} with the EEM calculated \emph{a priori} using Eqns~\eqref{eq:muprior} and \eqref{eq:sigmaprior}, Approximation~\ref{appr:aemsi} with the EEM calculated adaptively over the posterior distribution converging to Eqns~\eqref{eq:postmu} and \eqref{eq:postsigma}, and Approximation~\ref{appr:aemsd} with the EEM calculated adaptively over the posterior distribution. All cases used GCAM for the proposal with the target acceptance rate of $13\%$.
The acceptance rate in Step~\ref{step:ada2} of ADA, $\bar{\beta}$, and the IACT of the likelihood function are shown in Table~\ref{table:performance}. (We did not run Approximation~\ref{appr:local} for this model.)
\begin{table}[ht]
\caption{Performance summary for various approximations to the well test model. Shown are the second step acceptance rate $\bar{\beta}$, and IACT of the log-likelihood function.}
\label{table:performance}
\centering
\begin{tabular}{llllll}
\toprule
          & Approx. 1 & Approx. 2 & Approx. 2   & Approx. 4 & Standard MH \\
          &           & (prior)   & (posterior) &           &   \\
\midrule
$\bar{\beta}$   & $0.12$ & $0.31$ & $0.77$   & $0.93$   & $1$     \\
IACT            & -      & -      & $208$    & $153$    & $169$ \\
\bottomrule
\end{tabular}
\end{table}

Approximation~\ref{appr:coarse} (using the reduced model directly) only produces $\bar{\beta}=12 \%$. (This agrees closely with the equivalent method in~\cite{Efendiev}.) Approximation~\ref{appr:aemsi} with EEM built over the prior, as in Eqs~\eqref{eq:muprior} and \eqref{eq:sigmaprior}, increases the acceptance rate in step~\ref{step:ada2} of ADA to $\bar{\beta}=31 \%$.
However, both Approximation~\ref{appr:coarse} and Approximation~\ref{appr:aemsi} with the EEM constructed over the prior cannot produce a well mixed Markov chain, even after $2\times10^5$ iterations, so the IACT for the log-likelihood function could not be estimated, and is not reported for these cases. By using formula \eqref{eq:su_factor}, we see that the use of Approximation~\ref{appr:aemsi} (prior) only improves computational efficiency marginally, while the use of the simple Approximation~\ref{appr:coarse}, as in~\cite{Efendiev}, actually reduces computational efficiency.

Approximation~\ref{appr:aemsi} with the EEM calculated adaptively over the posterior distribution produces significantly better mixing, with an estimated $\bar{\beta}=77 \%$, and IACT of the log-likelihood function of  $208$. 
Approximation \ref{appr:aemsd}, with EEM calculated adaptively over the posterior distribution and with state-dependent reduced model \eqref{eq:rom_l}, further improves performance, achieving $\bar{\beta}=93 \%$, and the IACT of the log-likelihood function is $153$. 
Using formula \eqref{eq:su_factor}, we estimate the factor by which computational efficiency is improved for ADA with Approximation~\ref{appr:aemsi} (posterior) and Approximation~\ref{appr:aemsd} is about $4.3$ and $5.9$, respectively.
We also notice that the IACTs of the log-likelihood function suggest that ADA with Approximation~\ref{appr:aemsd} is more statistically efficient than the standard MH, which cannot be the case as discussed in Section~\ref{sec:ce}. This effect is probably finite-sampling error in the IACT estimate. 
However, this result suggests that the decrease of statistical efficiency may be negligible.

Assuming the convergence of $\bar{\beta}$ is in the asymptotic regime, the theorem in Section~\ref{sec:betabar} implies that estimates calculated using an MCMC with the approximate forward map only incur a relative error of $0.88$ for the typical case where the deterministic reduced model is used with the original likelihood function in Approximation~\ref{appr:coarse} and is reduced by a factor of more than $10$ to $0.07$ for Approximation~\ref{appr:aemsd}, for the same computational cost. Of course it is somewhat unrealistic to suggest that Approximation~\ref{appr:aemsd} may be used stand-alone as it uses the local correction to produce a state-dependent reduced model. However, once the approximate likelihood function is tuned in Approximation~\ref{appr:aemsi} (posterior), this reduced model can be used stand-alone and has a relative error in any posterior estimate of $0.23$, roughly $4$ times better than Approximation~\ref{appr:coarse}.

\begin{figure}[ht]
\centering
\includegraphics[width=0.8\textwidth]{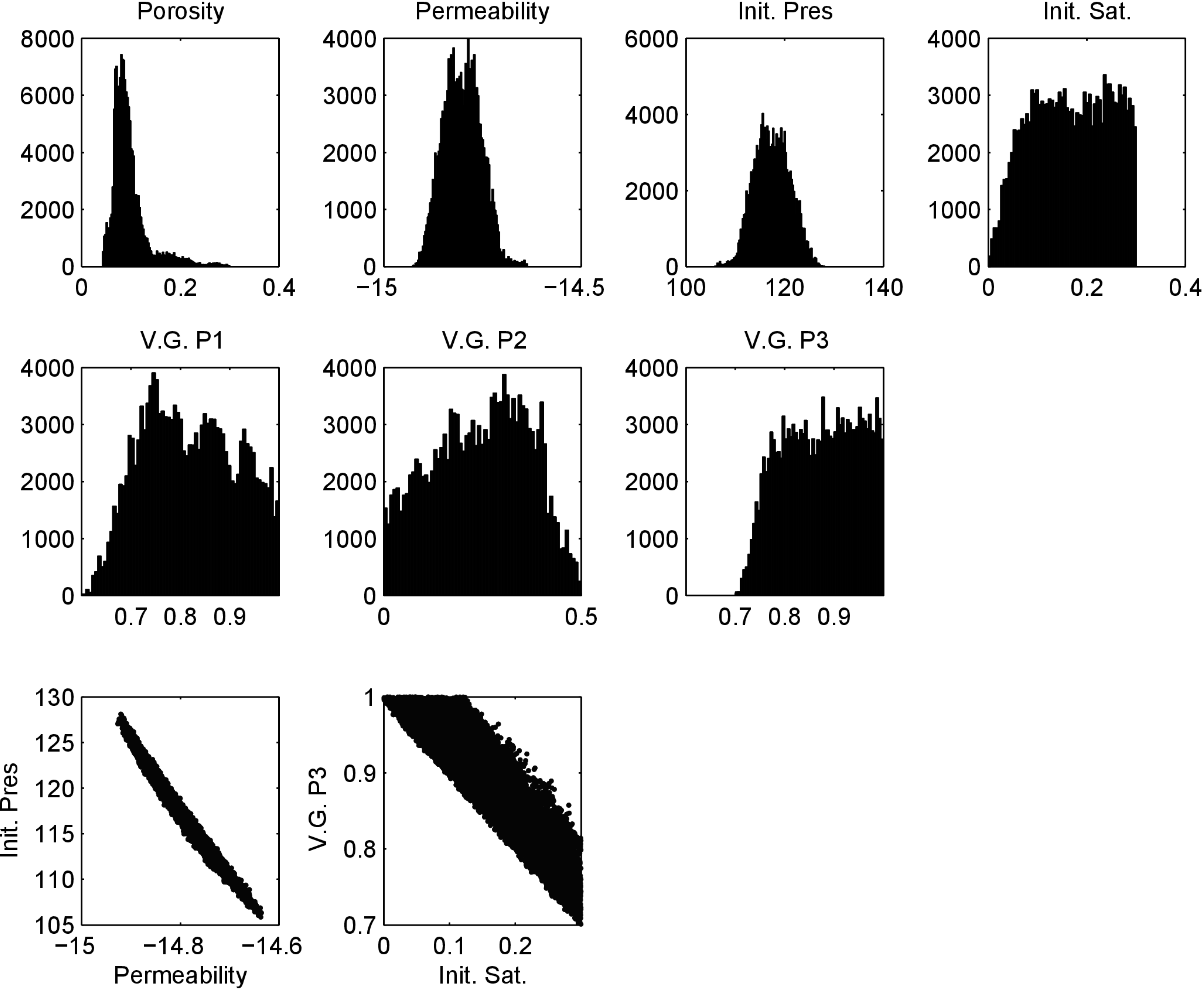}
\caption{Histograms of the marginal distributions and scatter plots between parameters, for the synthetic data set.}
\label{figure:s_hist}
\end{figure}

To indicate the nature of the posterior distribution, we show histograms of the marginal distributions in the first two rows of Fig.~\ref{figure:s_hist}. The parameter $\bx$ shows skewness in porosity and two of the hyperparameters of the van Genuchten-Mualem relative permeability model ($m$ and $S_\textrm{rl}$). The scatter plots between parameters show strong negative correlations between the permeability (on base $10$ logarithmic scale) and the initial pressure; see the left plot of last row of Fig.~\ref{figure:s_hist}. There is also a strong negative correlation between the initial saturation and one of the hyperparameters of the van Genuchten-Mualem ($S_\textrm{ls}$); see the right plot of last row of Fig.~\ref{figure:s_hist}. 

\subsection{Natural state modelling}
\label{sec:3d}

We now present an application of ADA using Approximation~\ref{appr:aemsd} to fitting a large-scale 3D  geothermal reservoir model using measured field data. 
We aim to infer the  permeability structure within the reservoir and the mass input at the bottom of the reservoir from temperature data measured from wells, and also to predict the size and shape of the hot plume of the reservoir.

\begin{figure}[ht]
\centering
\includegraphics[width=0.8\textwidth]{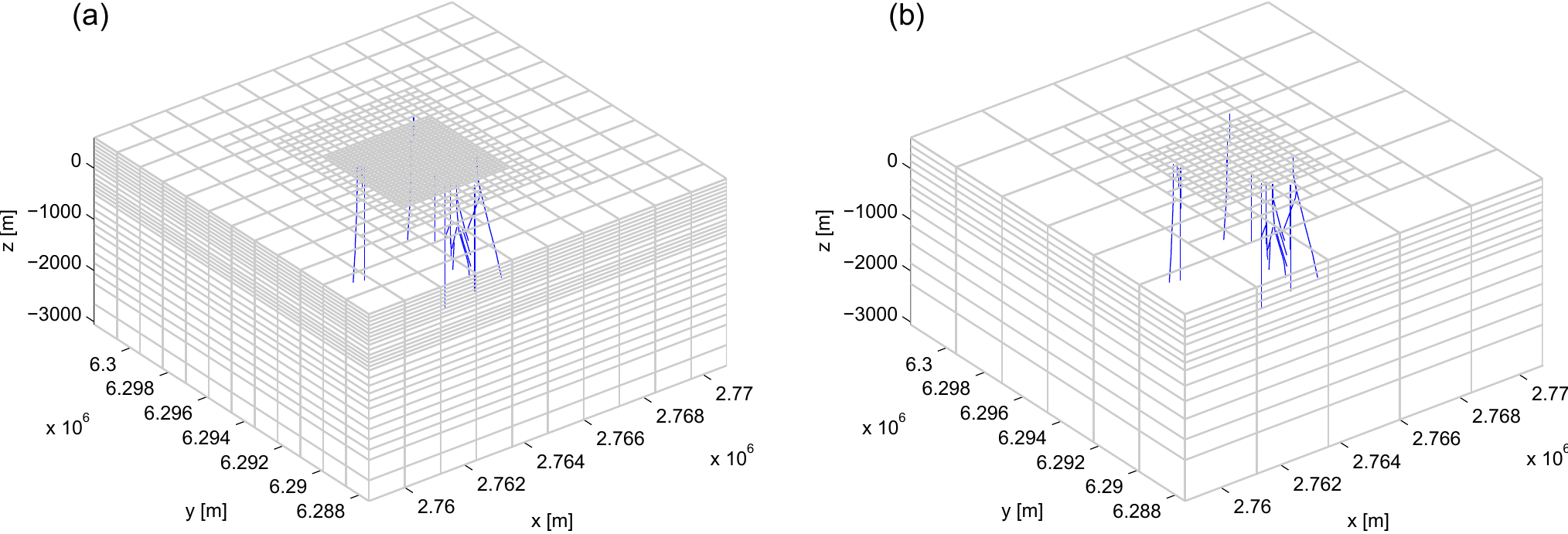}
\caption{The fine grid (left) and the coarse grid (right) used for natural state modelling.}
\label{figure:mokai}
\end{figure}

The 3D structure of the forward model has $26,005$ blocks, and is shown in Figure~\ref{figure:mokai} (a), where the blue lines in the middle of the grid show wells drilled into the reservoir.
The volume is $12.0$ km by $14.4$ km extending down to $3050$ meters below sea level. 
Relatively large blocks were used near the outside of the model and then were progressively refined near the wells to achieve a well-by-well allocation to the blocks. 
To speed up the computation a reduced model based on a coarse grid with $3,335$ blocks is constructed by combining adjacent blocks in the $x$, $y$ and $z$ directions; see Figure \ref{figure:mokai} (b). 
Each simulation of the forward model takes about $30$ to $50$ minutes CPU time on a DELL T3400 workstation, and about $1$ to $1.5$ minutes for the reduced model. Computing time for these models is sensitive to the input parameters. 

\begin{figure}[ht]
\centering
\includegraphics[width=1\textwidth]{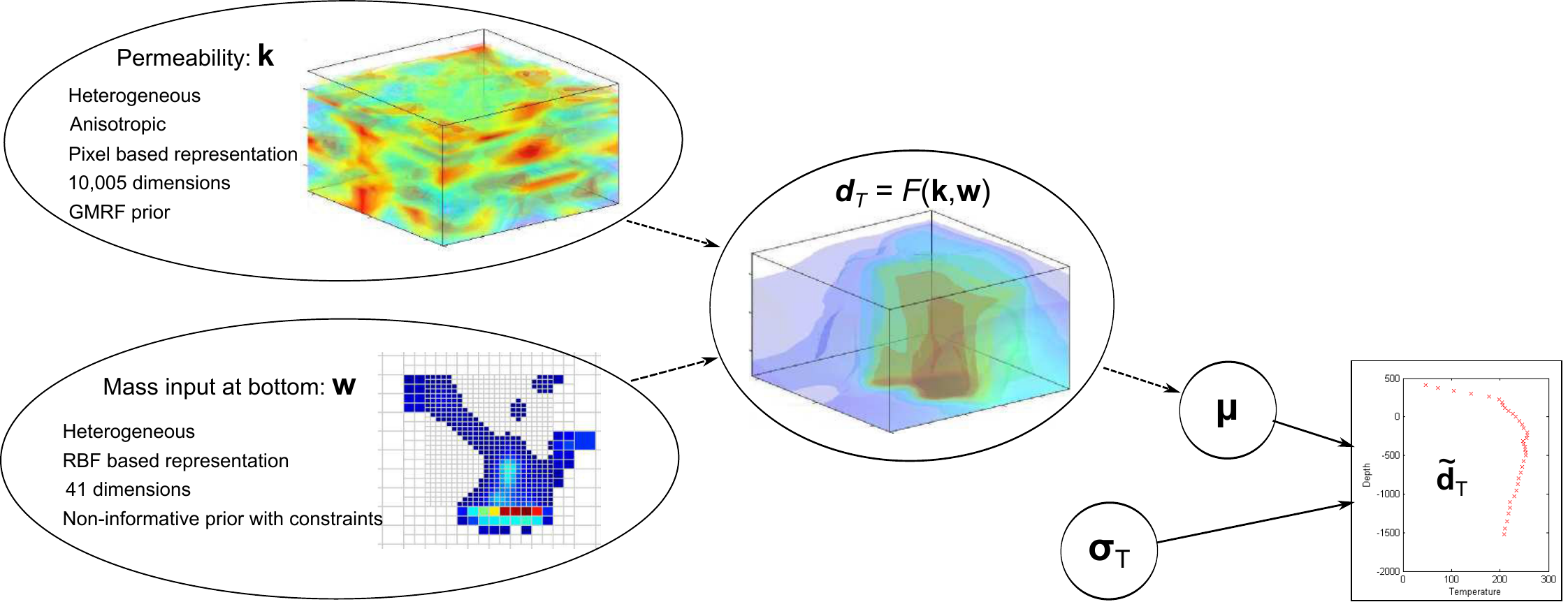}
\caption{DAG showing the hierarchical Bayesian model for calibration of the natural state geothermal reservoir.}
\label{figure:ns_3d}
\end{figure}

A DAG showing the hierarchical structure of this model is in Fig.~\ref{figure:ns_3d}.  State variables are the spatially distributed and heterogeneous permeabilities $\bf k$ and mass input from depth $\bf w$.
Permeabilities $\bf k$ are represented by a $10,005$ dimension voxel representation, with spatial correlation modelled by a Gaussian Markov random field prior~\eqref{eq:3d_prior}.
The mass input is modeled by a mid-level radial basis function expansion with the squared-exponential kernel function
\[
q_{\rm m} = \sum_{i=1}^{41} w_i \exp\left\{ \left( \frac{\| {\bx} - {\bx}_s\|}{r} \right)^2 \right\},
\]
with the $41$ dimensional weighting variable ${\bf w} = (w_1,\ldots,w_{41})^T$ associated with pre-specified control points ${\bx}_s, s = 1,\ldots,41$ determined from previous geophysical exploration. The constraints ${\bf w} > 0$ and $\sum {\bf w} = constant$ ensure that the mass input is positive and has a fixed total amount.

Simulating the forward model $F$ with parameters $(\bf k, w)$ produces model outputs of the temperatures ${\bd}_\textrm{T} = F(\bf k, w)$. Empirical estimation of the noise vector~\cite{wrr_cfo_2011} suggests an i.i.d. Gaussian distribution with standard deviation $\sigma_\textrm{T} = 7.5^{\circ}$C to be used in the likelihood function. This yields the posterior distribution
\begin{eqnarray}
\pi({\bf k, w}|\tilde{\bd}_\textrm{T}) & \propto & \exp\left[-\frac{1}{2{\sigma_\textrm{T}}^2}  \left\{F({\bf k, w}) - \tilde{\bd}_\textrm{T}\right\}^T  \left\{F({\bf k, w}) - \tilde{\bd}_\textrm{T}\right\}\right] \times \label{eq:3d_llkd} \\
& & \exp\left[ - \tau \sum_{i \sim j} \left\{\log_{10}(k_i) - \log_{10}(k_j)\right\}^2 \right] \times \chi({\bf k}) \label{eq:3d_prior} \\
\rm subject \ to & & \rm \sum {\bf w} = constant \ and \ {\bf w} > 0, \nonumber
\label{eq:3d_post}
\end{eqnarray}
where $\tau$ is a hyperparameter that controls smoothness, and $\chi({\bf k})$ is the indicator function for the prior bounds modelled in~\cite{wrr_cfo_2011}.
 
In this example, all approximations without the posterior tuned approximate likelihood and state-dependent local correction produce MCMCs that do not mix as the reduced model is simply too approximate. Using Approximation~\ref{appr:aemsd} we are able to sample the posterior distribution for about $11,200$ iterations in $40$ days, and ADA achieves about $\bar{\beta}=74\%$ acceptance rate in the second accept-reject step. The estimated speed-up in computational efficiency is by a factor of $7.7$, and the estimated IACT of the log-likelihood function is about $5.6$. This is only a rough estimate because the chain has not been running long enough, however all the samples show a good fit to measured data; see~\cite{CuiFoxO'Sullivan2019} for further details.

\subsection{Electrical Capacitance Tomography}
\label{sec:ect}
\begin{figure}[ht]
\centering
\includegraphics[width=0.3\textwidth]{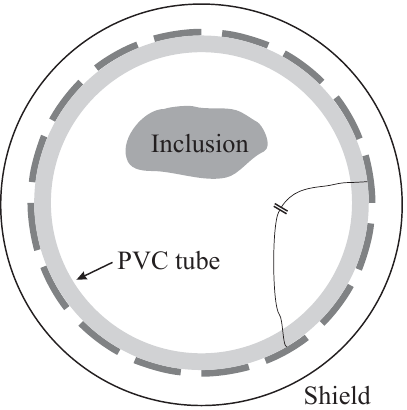} 
\hspace{0.2\textwidth}
\includegraphics[width=0.3\textwidth]{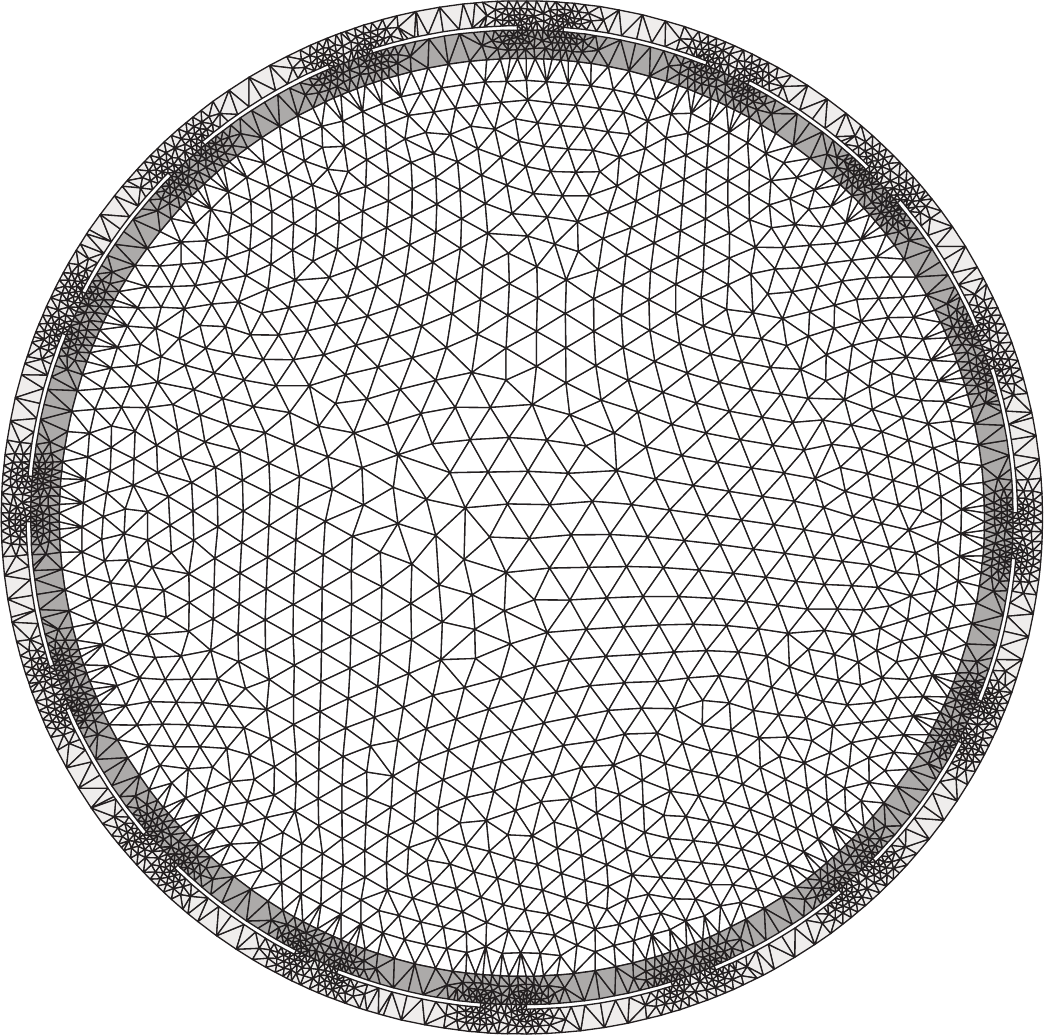}
\caption{Left: Scheme of an ECT sensor. Right: A finite element mesh used for ECT.}
\label{fig:ect}
\end{figure}

Electrical capacitance tomography (ECT) is an inverse problem which uses measurements of the inter-electrode capacitances to determine the spatially dependent dielectric permittivity distribution in a region of interest $\Omega_\text{ROI}$.
Figure~\ref{fig:ect} (left) 
depicts a typical scheme for a 2D ECT system suitable for process tomography. A number 
of electrodes are mounted on the exterior of a process pipe (typically a PVC tube).
Typically an AC voltage is applied to one of the electrodes
and the displacement currents on the other electrodes are measured; see~\cite{WatzenigFox} for further details of measurements.

When electrode $i$ is held at potential $V_0$ and all others are held at virtual earth, the potential $V_i$ in the whole region $\Omega$  satisfies the Dirichlet boundary value problem (BVP),
\begin{eqnarray}
  \label{eqn:forward}
  \nonumber \nabla\cdot(\varepsilon \nabla V_i)=0,
     &\qquad \mbox{in $\Omega$},\\
  V_i|_{\Gamma_i}=V_0,\\
  \nonumber V_i|_{\Gamma_j}= 0, &\qquad j\neq i, \\
  \nonumber V_i|_{\partial\Omega}=0,
\end{eqnarray}
where both the permittivity $\varepsilon(r)$ and potential
$V_i(r)$ depend on position $r\in\Omega$. $\Gamma_j$ denotes the surface of electrode $j$. After solving BVP~\eqref{eqn:forward} for the specified boundary conditions, the capacitance $C_{i,j}$ between electrode $i$ and electrode $j$ can
be computed by
\begin{equation}
  C_{i,j}=-\frac{1}{V_0}\oint\limits_{\Gamma_{j}}\varepsilon\frac{\partial
  u_i}{\partial\mathbf{n}}\mbox{d}r
  \label{eq:charge}
\end{equation}
where $\mathbf{n}$ is the inward normal vector. 
The forward map is defined by the map $F:\varepsilon|_{\Omega_\text{ROI}}\mapsto C$.

We simulate this forward map using a finite element method (FEM) discretization of the region, using continuous piecewise linear functions on triangles. In this section we investigate the use of a coarse-mesh reduced model for the region of interest, and investigate how coarse a reduced model can be while still being accurate. In particular, our interest is in tuning a very coarse surrogate that may then be used standalone for inference embedded in the sensor. Accordingly, we consider Approximations~\ref{appr:coarse} and \ref{appr:aemsi} that do not use the local correction (since that requires also simulating the accurate model in DA). 

Typically offset capacitances between the electrodes are in the range of pF, whereas the changes caused by inclusions
are in the range of fF. That is, signal deviation is only a fraction of the offset value so calibration of measurements is necessary and ECT is a differential imaging method. Further properties of the forward map $F:\varepsilon\mapsto C$ are given in~\cite{WatzenigFox}, while details of deterministic (least-squares) and Bayesian calibration are presented in~\cite{NeumayerPhD}.

Figure~\ref{fig:ect} (right) shows an ‘unstructured’  FEM mesh with about $6000$ elements,
used for solving the BVP~\eqref{eqn:forward} so that discretization error is smaller than a typical signal-to-noise-ratio (SNR) of  1000:1~\cite{SchwarzlMSc}. The discretized area includes the insulating
pipe (dark grey) the region outside the pipe with electrode inset
(light grey) and the region of interest $\Omega_\text{ROI}$ inside the pipe. This mesh has smaller elements around the electrode ends to give accurate representation of
rapid changes in fields, and with larger elements in $\Omega_\text{ROI}$ towards the
centre of the pipe where decreased resolution of ECT does not warrant finer division of the permittivity~\cite{WatzenigFox}.

The `charge map' approach is a method whereby simulation of the forward map may be computed over the FEM mesh in $\Omega_\text{ROI}$, only, to avoid solving the full FEM system. This approach computes the charges on
the electrodes as a function of the potential distribution on $\partial\Omega_\text{ROI}$, which
is the interior boundary of the tube. More precisely, charges on electrodes are split into a constant part and a part
which depends linearly on $V|_{\partial\Omega_\text{ROI}}$. Thus the forward map may be reduced to solving for the potential in the domain $\Omega_\text{ROI}$, essentially by a Woodbury formula applied to the Schur compliment that reduces solving the full FEM system to solving only systems of the size of the smaller FEM matrix on $\Omega_\text{ROI}$. Details may be found in~\cite{NeumayerPhD}. 

\begin{figure}[ht]
\centering
\includegraphics[width=0.3\textwidth]{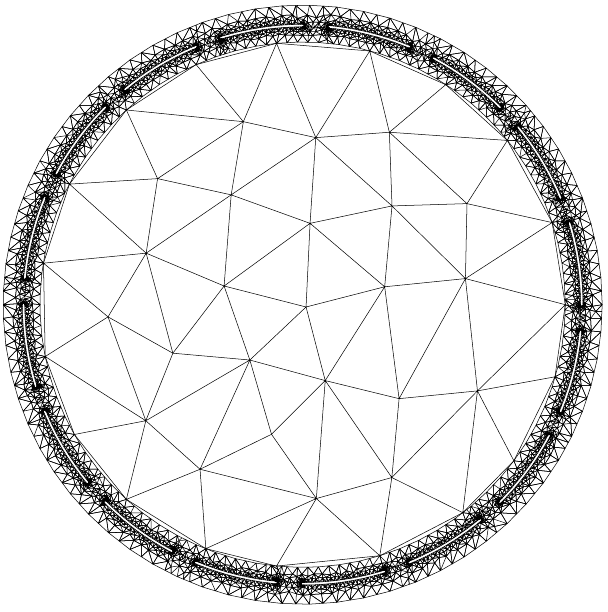} 
\hspace{0.2\textwidth}
\includegraphics[width=0.3\textwidth]{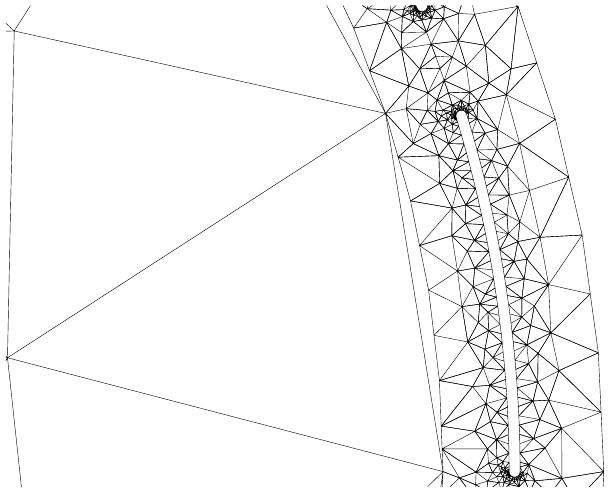}
\caption{Left: Finite element mesh using just $76$ elements in $\Omega_\text{ROI}$. Right: Closeup view of the tube internal boundary showing the gap between meshes caused by the coarse mesh.}
\label{fig:ect76}
\end{figure}

We denote the FEM solver by $P:\beps\mapsto\bd$, and will consider three meshes for $\Omega_\text{ROI}$ having $2000$, $600$ and $76$ elements, denoted $P_{2000}$, $P_{600}$ and $P_{76}$, respectively. Compute times for these forward maps are in the ratio $260:48:13$.  The coarsest mesh defining $P_{76}$ is shown in Fig.~\ref{fig:ect76} (left). 
Simulated data was generated using $P_{2000}$ with noise added, and treated as physical measurements $\tilde{\bd}$ with a signal-to-noise ratio (SNR) of 1000:1. Using a finer mesh for simulation than for reconstruction avoids the most obvious `inverse crimes'~\cite{KaipioSomersalo2007}. In particular, we generated calibration data $\tilde{\bd}_0$ and $\tilde{\bd}_1$ corresponding to an empty pipe $\beps_0$ and pipe with inclusions $\beps_1$, respectively. Reconstruction is performed with $P_{600}$ as the accurate model and $P_{76}$ as the reduced model. As is required in practical ECT, each of the FEM models was offset-gain corrected before using for reconstruction. We used simple least-squares calibration in this experiment, i.e. we set 
\begin{equation}
 F^*_k(\beps) = \brho_k P_k(\beps) + \bc_k,\quad k=600,76,
 \label{eq:lscalib}
\end{equation}
where
\begin{equation}
 \brho_k = \frac{\tilde{\bd}_1-\tilde{\bd}_0}{F^*_k(\beps_1)-F^*_k(\beps_0)}
 \quad\mbox{and}\quad
 \bc_k = \frac{P_k(\beps_0))\tilde{\bd}_1-P_k(\beps_1))\tilde{\bd}_0}{F^*_k(\beps_1)-F^*_k(\beps_0)}
\end{equation}

Using $F^*_{600}$ within MH produces a useful reconstruction of the unknown permittivity in $\Omega_\text{ROI}$, with quantified uncertainties~\cite{NeumayerPhD}. This resolution mesh is often used in embedded regularization-based inversion as a best trade-off between accuracy and cost. 

The very coarse mesh used in $F^*_{76}$ has too few elements to allow reconstruction, without modification. Indeed, the forward map for ECT has an effective rank of about $110$ at the SNR used~\cite {WatzenigFox}, so $76$ elements is not even sufficient to represent the range of the actual forward map. A further problem with the mesh in $F^*_{76}$  is the gap between the coarse inner mesh and the fine outer mesh; see Fig.~\ref{fig:ect76} (right).  At `free nodes', where the two meshes do not connect, we used a simple linear interpolation of potentials at `connected nodes' where the two meshes coincide.  Using $F^*_{76}$ without randomizing within DA, as in Approximation~\ref{appr:coarse} ($F^*_{600}$ is the accurate forward map), produces an MCMC that that cannot interpret data, and does not converge to the true posterior distribution over useful time scales. 

We also randomized $F^*_{76}$ via Approximation~\ref{appr:aemsi} with the EEM calculated adaptively over the posterior distribution. Combined with the gain term in Eqn~\eqref{eq:lscalib} fit by least-squares, this gives an approximate like function with the form of Eqn~\eqref{eq:likestar1}. This gave significantly better mixing with DA converging and providing the same estimates as the MH MCMC running $F^*_{600}$. The value of $\bar{\beta}=53\%$ indicates that  $F^*_{76}$ with the modified likelihood function is an acceptable approximation, and using formula \eqref{eq:su_factor} we estimate that this use of $F^*_{76}$ gives a speed-up in computational efficiency by a factor of $3.0$. This is quite remarkable when we consider that $F^*_{76}$ used directly is not accurate enough to allow useful imaging.

However, the low value of $\bar{\beta}=0.53$ indicates that this corrected approximation is not good enough to be used as a standalone approximation, without correction in DA. Undoubtedly, using the local correction in Approximation~\ref{appr:aemsd} would further improve $\bar{\beta}$ in DA, but would defeat the purpose of this example of finding a surrogate to replace the expensive calculation for embedded inference. We mentioned the disconnection between coarse and fine meshes used in $F^*_{76}$ and we believe that this is the main residual source of error. This is particularly evident when imaging high-contrast permittivity inclusions where non-linearity of the forward map is more prominent and the error produced by disconnected meshes is severe; the results presented above are for low-contrast inclusions. We conjecture that an improvement in our simple linear-interpolation of nodal values would improve the approximation, though we have not tested specific fixes.

\section{Discussion and a Comparison}
\label{sec:dis}
We considered sample-based uncertainty quantification for inverse problems within the Bayesian formulation. Our primary contribution has been to show that randomizing a deterministic reduced model, or, equivalently, using a modified likelihood function, can improve the resulting approximation to the posterior distribution. This is evident when the improved approximation is used within the the DA algorithm~\cite{ChristenFox2005} and leads to an increase in the second-step acceptance rate, with the best case giving rates close to 1. Since the randomization requires negligible extra computation in evaluating the modified likelihood function, any increase in statistical efficiency translates directly to an increase in computational efficiency. This leads to lower compute cost required to evaluate estimates to within a desired accuracy. Tuning of the randomization was performed within the adaptive delayed acceptance (ADA) algorithm that builds stochastic improvements to reduced models, at negligible increase in computational cost over standard DA. Quality of approximations was measured by the second-step acceptance rate in DA, that relates directly to quality of estimates when using only the reduced model and modified likelihood function. 

The observation that randomization can improve a `best' deterministic reduced model may appear counter intuitive, however the quantitative results presented here are unequivocal; besides, randomizing is equivalent to using a modified likelihood function tuned to best approximate the target posterior distribution. We built the randomized reduced map by evaluating posterior statistics of the reduced model, following existing models for model error in~\cite{KaipioSomersalo2004,KennedyO'Hagan2001}. This was possible by operating in an adaptive version of the delayed acceptance algorithm, that also enabled a zeroth-order local correction to produce a state-dependent reduced model. In the examples in geothermal reservoir calibration, we found that using a state-dependent approximation and posterior EEM is critical to improving computational efficiency. In contrast, not using the local correction and not using the EEM or estimating it over the prior distribution did not lead to appreciable improvement in computational efficiency, or actually decreased it.

We have not proved that randomizing a reduced model, or the particular randomization used here, necessarily increases computational efficiency, except for the observation that it is not necessarily worse. Indeed, we would not be surprised if there are inverse problems and approximations for which the methods here offer no improvement. However, we have demonstrated that these ideas can lead to significant improvement in computational efficiency of sample-based inference in practical and large-scale inverse problems, and we have quantified both the improvement in computational efficiency and the quality of approximations in the computed examples.

Quality of the approximate target distribution induced by an reduced forward model may be measured by the second-step acceptance rate $\bar{\beta}$ in DA, as shown in Section~\ref{sec:betabar}. Further, when $\bar{\beta}\approxeq 1$ the accurate model can be discarded. We investigated this possibility in a computed example of ECT, by posterior tuning of a modified likelihood function for use with a very coarse reduced model that used only $76$ elements in the region of interest. While this randomized reduced model gave an improvement in computational efficiency when used in DA, the approximation was not good enough to be used standalone for inference. We expect that further \textit{ad hoc} improvements handling disconnected meshes will provide better approximations.

We present a proof of ergodicity for the ADA algorithm using any of the approximations in Section~\ref{sec:approx}, by drawing on the simplified conditions for adaptive sampling in~\cite{RobertsRosenthal2007}. In this way we have produced a happy hunting ground for practitioners to discover and utilize other approximations. Any approximation that is cheap and accurate is a valid starting point for the local corrections and modified likelihood functions that we presented, and gives guaranteed convergence, which means that application-specific intuition, or even just guesses, may be leveraged to produce more efficient algorithms for sample-based inference. Quality of an approximation and modification of the likelihood function may be quantified, as described above.

\cf{
Contributions made in this paper may be emphasized by comparing to the iterative updating of the EEM presented in ~\cite{calvetti2018iterative}, that has broadly similar aims of improving upon the EEM to improve inference in inverse problems when using an approximate forward map that reduces computational cost. 
Calvetti \emph{et al.}~\cite{calvetti2018iterative} improved on the \textit{a priori} EEM of~\cite{KaipioSomersalo2004} using an iterative updating, noting that since the \textit{a priori} EEM produces a cheap-to-calculate approximate posterior distribution with more accurate conditional posterior mode, the estimation of statistics in the EEM can be performed again using samples from this improved approximate distribution, and then iterate repeatedly.
This generates a fixed linear iteration on the space of probability distributions, so convergence is geometric, at best, though convergence is not guaranteed under all conditions. In the most accurate scheme in~\cite{calvetti2018iterative} that uses full representation of distributions using a particle method, that would be computationally expensive in practice, even if a limit exists it is not guaranteed to be the best approximation to the true posterior. Indeed, since the algorithm in~\cite{calvetti2018iterative} bounds computational cost by only using evaluations of $F$ at states drawn from the original prior, it seems likely that the iteratively updated EEM will remain an approximation to the prior distribution, producing inaccurate posterior inference in inverse problems with highly informative data. In contrast, ergodicity of the ADA algorithm guarantees convergence to posterior estimates of the EEM, and convergent posterior inference; see Appendix~\ref{sec:ergo}. Since multiple evaluations of the true map $F$ are required for tuning the EEM, including the updated EEM~\cite{calvetti2018iterative}, and the resulting EEM only gives an approximation to the true posterior, it is not clear what computational efficiency is achieved by the iterative updating in~\cite{calvetti2018iterative}. A suitable comparison of efficiencies can be made using the computed EIT example in~\cite{calvetti2018iterative} and the related ECT inverse problem in Section~\ref{sec:ect}. These computed examples have similar forward maps and report a similar number of posterior samples being required for evaluating estimates, i.e., $5000$ and $4000$ respectively. Tuning of the EEM in~\cite{calvetti2018iterative} uses $1500$ draws from the prior, hence $1500$ evaluations of $F$. The speedup by a factor of $3.0$ reported in our ECT example implies that $4000/3=1300$ evaluations of the exact map $F$ are required for the MCMC; hence a comparable number of full function evaluations are required in the two cases. However, the EIT example in~\cite{calvetti2018iterative} only produces estimates over the approximated posterior distribution\footnote{Even though~\cite{calvetti2018iterative} discusses Bayesian posterior inference, the paper only reports point estimates evaluated as conditional posterior modes for fixed hyperparameters (a.k.a., regularized inverses) that are not well defined as a Bayesian posterior statistic~\cite{wiki:MAP}.} whereas the ADA algorithm allows evaluation of expectations over the correct posterior distribution. Further, the more general approximations and optimal tuning performed with ADA make feasible more coarse, hence cheap, approximations to be used, and so the cost of evaluating the approximate forward map in the ECT example in Section~\ref{sec:ect} is reduced in ADA compared to the EIT example in~\cite{calvetti2018iterative}. On the basis of these estimates, the iterative updating of the EEM in~\cite{calvetti2018iterative} appears to produce a less efficient and less accurate algorithm for large-scale problems, compared to ADA\footnote{This agrees with the observation in~\cite{wrr_cfo_2011} that ADA produces posterior estimates of \emph{any} quantity within the computational cost of just tuning the EEM over the prior distribution.}. Both the EIT example in~\cite{calvetti2018iterative} and our ECT example use a fixed approximate forward map. As noted above, ADA also allows the use of a locally-corrected state-dependent approximation that further improves computational efficiency, see Section~\ref{sec:statedepend}, that could further improve computational efficiency in our ECT example, as noted in Section~\ref{sec:ect}; no such efficiency is available with the updated EEM in~\cite{calvetti2018iterative}. 

}

\appendix

\section{Ergodicity of ADA}
\label{sec:ergo}
We follow the notation in \cite{RobertsRosenthal2007} to formalize ADA. In particular, we index distributions by adaptation indices, rather than iteration number, as in Alg.~\ref{alg:ada}, since the former provides a unique notation for functions. To simplify notation in this section, let $\pi(\cdot) = \pi_\text{post}(\cdot|\tilde{\bd})$ denote the exact posterior distribution.

Suppose $\pi(\cdot)$ is a fixed target distribution, defined on state space $\mathcal{X}$ with $\sigma$-algebra $\mathcal{B(X)}$.  
Let $\{K_\bgamma\}_{\bgamma\in\mathcal{Y}}$ be a family of Markov chain transition kernels (associated with MH) on $\mathcal{X}$, and suppose that for all $\bgamma \in \mathcal{Y}$, $\pi(\cdot)$ is the unique stationary distribution. 
Let $\{\pi_{\bxi,{\bx}}^*(\cdot)\}_{\bxi\in\mathcal{E}}$ be a family of state-dependent approximations to the exact target distribution $\pi(\cdot)$ for all $\bxi \in \mathcal{E}$. 

The adaptation indices $\bgamma$ and $\bxi$ are associated with adaptation of the proposal and approximate target, respectively.  
At each step $n$, ADA updates $\bgamma$ and $\bxi$ by a $\mathcal{Y}$-valued random variable $\boldsymbol\Gamma_n$ and a $\mathcal{E}$-valued random variable $\boldsymbol\Xi_n$, respectively. The transition kernel of ADA is denoted by $\{K_{\bgamma,\bxi}\}_{\bgamma\in\mathcal{Y}, \bxi\in\mathcal{E}}$. 

We prove the following theorem, that ergodicity of ADA can be guaranteed by imposing certain regularity conditions.

\begin{theo}
\label{theo3}
Consider an ADA algorithm, with target distribution $\pi(\cdot)$ defined on a state space $\mathcal{X}$, with $\mathcal{Y}$-valued proposal adaptation index and $\mathcal{E}$-valued approximation adaptation index. 

Suppose that for each $\bgamma \in \mathcal{Y}$, $K_\bgamma$ is the kernel of a MH algorithm targeting $\pi$ with proposal kernel $Q_\bgamma({\bx},\dd{\by}) = q_\bgamma(\by|\bx) \lambda(\dd{\by})$ having a density $q_\bgamma(\cdot|\bx)$ with respect to some finite reference measure $\lambda(\cdot)$, with corresponding density $h$ for $\pi(\cdot)$ so that $\pi(\dd{\by}) = h({\by})\lambda(\dd{\by})$. 
Similarly, for each $\bxi \in \mathcal{E}$, the state-dependent approximation $\pi^*_{\bxi,\bx}(\cdot)$ has density $h^*_{\bxi,\bx}(\cdot)$ such that, $\pi^*_{\bxi,\bx}(\dd{\by}) = h^*_{\bxi,\bx}({\by})\lambda(\dd{\by})$. 
Let $K_{\bgamma,\bxi}$ be the transition kernel of the corresponding ADA algorithm using the approximation $\pi_{\bxi,{\bx}}^*(\cdot)$, proposal $q_\bgamma$, and targeting $\pi(\cdot)$. Suppose further that the following conditions hold:

\begin{enumerate}
  \item The spaces $\mathcal{X}$, $\mathcal{Y}$, and $\mathcal{E}$ are compact. \label{cond:a}
  \item Each transition kernel $K_\bgamma$ is ergodic for $\pi(\cdot)$. \label{cond:b}
  \item For all $\bgamma \in\mathcal{Y}$, $q_\bgamma(\cdot,\cdot)$ is uniformly bounded, and the mapping $({\bx},{\by},\bgamma) \mapsto q_\bgamma(\by|\bx)$ is continuous. \label{cond:c}
  \item The proposal distribution satisfies diminishing adaptation, that is,\\ $\lim_{n\rightarrow\infty}\sup_{\bx}\|Q_{\boldsymbol\Gamma_{n+1}}({\bx},\cdot)-Q_{\boldsymbol\Gamma_{n}}({\bx},\cdot)\|_{TV} = 0$ in probability, where $\|\mu(\cdot)-\nu(\cdot)\|_{TV} = \sup_{{\bf A}\in\mathcal{B}(\mathcal{X})}\|\mu({\bf A})-\nu({\bf A})\|$ is the total variational norm. \label{cond:d}
  \item The mapping  $({\bx},{\by},\bxi) \mapsto \log h^*_{\bxi,\bx}({\by})$ is continuous. \label{cond:g}
  \item The approximation adaptation index satisfies diminishing adaptation, that is, $\lim_{n\rightarrow\infty} \| \boldsymbol\Xi_{n+1}-\boldsymbol\Xi_{n} \| = 0$ in probability.  \label{cond:h}
\end{enumerate}

Then ADA is ergodic for $\pi(\cdot)$.
\end{theo}

The regularity assumptions in Theorem~\ref{theo3} may look daunting, but are actually not restrictive for many practical applications. 
Condition~\ref{cond:a}, that parameter space and the adaptation spaces are compact, is often a consequence of physical bounds on the parameters and bounded model outputs. In practical computation, one could argue that this assumption always holds as computers are finite dimensional, though one does not want to explore the full range of numerical representations if the algorithm is to be efficient! 
Conditions~\ref{cond:b}, \ref{cond:c}, and \ref{cond:d} are conditions on the proposal distribution, and depend on the choice of proposal and adaptation that is used. These conditions can be satisfied by making suitable choices. 
Condition~\ref{cond:g} is satisfied by the forward model and its reduced model in most inverse problems; Indeed, the more ill-posed is the inverse problem, the more well-posed is the forward model and the higher the order of continuity satisfied by the forward model. Finite-dimensional reduced models are continuous because stiffness matrices are not singular when the reduced model is well posed.
Condition~\ref{cond:h}, of diminishing adaptation, follows when the adaptation is to some fixed property of the posterior distribution, as with the posterior statistics of the EEM that we use to parametrize the randomizing distribution.

\noindent \textbf{Proof:}
We prove Theorem~\ref{theo3} by establishing the conditions of Theorem 1 of \cite{RobertsRosenthal2007} for the composite adaptation index $\left(\bgamma,\bxi\right)$ and proposal $q^*_{\bgamma,\bxi}({\bx},\cdot)$. 
First note that, since $\mathcal{X}$, $\mathcal{Y}$, and $\mathcal{E}$ are compact, all product spaces are compact in the product topology.

By Theorem 1 of \cite{ChristenFox2005}, conditions~\eqref{cond:a}, \eqref{cond:b}, and  \eqref{cond:g} imply that, for all $(\bgamma,\bxi) \in\mathcal{Y}\times\mathcal{E}$ the transition kernel $K_{\bgamma,\bxi}$ is ergodic for $\pi(\cdot)$. 

The effective proposal in step~\ref{step:ada2} of ADA has density
\begin{equation*}
q^*_{\bgamma,\bxi}({\bx},{\by}) = \alpha_{\bgamma,\bxi}({\bx},{\by}) q_\bgamma(\by|\bx) + \{1-r_{\bgamma,\bxi}({\bx})\}\delta_{\bx}({\by}),
\end{equation*}
where 
\begin{equation*}
\alpha_{\bgamma,\bxi}({\bx}, {\by}) = \min\left\{ 1, \frac{h^*_{\bxi,{\bx}}({\by}) q_\bgamma({\by} ,{\bx})}{h^*_{\bxi,{\bx}}({\bx}) q_\bgamma(\by|\bx)}\right\} ,
\end{equation*}
and $r_{\bgamma,\bxi}({\bx})$ is the probability of accepting a proposal from ${\bx}$ in step~\ref{step:ada1} of ADA, given by
\begin{equation*}
r_{\bgamma,\bxi}({\bx}) = \int_{\mathcal{X}} \alpha_{\bgamma,\bxi}({\bx}, {\by})q_\bgamma(\by|\bx) \lambda(\dd{\by}) .
\end{equation*}
It follows from conditions \eqref{cond:c} and (\ref{cond:g}) that  
$({\bx},{\by},\bxi, \bgamma)\mapsto\alpha_{\bgamma,\bxi}({\bx}, {\by})q_\bgamma(\by|\bx)$ is continuous, and that $({\bx},\bgamma, \bxi)\mapsto r_{\bgamma,\bxi}({\bx})$ is continuous as in Corollary 5 of  \cite{RobertsRosenthal2007}\footnote{Note that continuity in ${\by}$ is required in condition \eqref{cond:c}; the conditions in Corollary 5 of  \cite{RobertsRosenthal2007} are not quite sufficient for general proposal distributions.}.

Hence the probability of accepting a proposal from ${\bx}$ in both steps~\ref{step:ada1} and \ref{step:ada2} of ADA is
\begin{equation*}
\rho_{\bgamma,\bxi}({\bx}) = \int_{\mathcal{X}} \beta_{\bgamma,\bxi}({\bx}, {\by})\alpha_{\bgamma,\bxi}({\bx}, {\by})q_\bgamma(\by|\bx) \lambda(\dd{\by})
\end{equation*}
where the acceptance probability in step~\ref{step:ada2} of ADA,
\begin{equation*}
\beta_{\bgamma,\bxi}({\bx}, {\by}) = \min\left\{ 1, \frac{h({\by}) \alpha_{\bgamma,\bxi}({\by}, {\bx})q_\bgamma({\by} ,{\bx})} {h({\bx}) \alpha_{\bgamma,\bxi}({\bx}, {\by})q_\bgamma(\by|\bx)} \right\},
\end{equation*}
is jointly continuous in ${\bx}$, ${\by}$, $\bgamma$ and $\bxi$. It follows, as in Corollary 5 of  \cite{RobertsRosenthal2007}, that $K_{\bgamma,\bxi}$ satisfies the simultaneous uniform ergodicity condition in Theorem 1 of \cite{RobertsRosenthal2007}.

Diminishing adaptation of the overall transition kernel $K_{\bgamma,\bxi}$ follows, as in Lemma 4.21 in~\cite{LRR2013}, from diminishing adaptation of the proposal $q^*_{\bgamma,\bxi}({\bx},{\by})$, which can be established using the triangle inequality treating adaptation indices $\bgamma$ and $\bxi$ in separate steps. Diminishing adaptation of $q^*_{\bgamma,\bxi}({\bx},{\by})$ with respect to adaptation in $\bgamma$ follows directly from condition~\ref{cond:c}, again as Lemma 4.21 in~\cite{LRR2013}. Diminishing adaptation of  $q^*_{\bgamma,\bxi}({\bx},{\by})$ with respect to adaptation in $\bxi$ may be established using the inequality
\begin{align*}
 \|q^*_{\bgamma,\bxi_{n+1}}({\bx},\cdot)&-q^*_{\bgamma,\bxi_n}({\bx},\cdot)\|_{TV} \\
 \leq & 2\int_{\mathcal{X}}\min\left\{ q_\bgamma(\by|\bx), 
 \left|\frac{h^*_{\bgamma,\bxi_{n+1}}({\by})}{h^*_{\bgamma,\bxi_{n+1}}({\bx})}-\frac{h^*_{\bgamma,\bxi_{n}}({\by})}{h^*_{\bgamma,\bxi_{n}}({\bx})} \right|  q_\bgamma({\by} ,{\bx}) \right\} \lambda(\dd {\by}).
\end{align*}
From conditions~\ref{cond:a}, \ref{cond:c} and \ref{cond:g} it follows  that the RHS $\rightarrow 0$, uniformly, as $\|\bxi_{n+1}-\bxi_{n}\|\rightarrow 0$. Diminishing adaptation of of $q^*_{\bgamma,\bxi}({\bx},{\by})$ with respect to adaptation in $\bxi$ follows from Condition~\ref{cond:h}. 

Thus the conditions in Theorem 1 of \cite{RobertsRosenthal2007} are satisfied, and the result follows.
\quad$\qed$

We now use Theorem~\ref{theo3} to establish ergodicity of ADA when using the approximation schemes in Section~\ref{sec:approx}.

\begin{coro}
\label{coro1}
Suppose that the forward map $F(\cdot)$ and its reduced model $F^*(\cdot)$ are continuous functions. Then the ADA Alg.~\ref{alg:ada} with  an adaptive proposal that satisfies diminishing adaptation in Alg.~\ref{alg:adapt}, using any of the Approximations~\ref{appr:coarse} to \ref{appr:aemsd} in Section~\ref{sec:approx}, is ergodic for $\pi(\cdot)$. 
\end{coro}
\noindent \textbf{Proof:}
This result follows from compactness of the space of possible $\boldsymbol\mu_{\bB,n}$  and $\mathbf{\Sigma}_{\bB,n}$, and that the proposal satisfies diminishing adaptation. See~\cite{CuiFoxO'Sullivan2019} for details. 
\quad$\qed$

\begin{acknowledgements}
This paper is a written and expanded version of the keynote presentation by CF at the FrontUQ18 workshop in Pavia, 5-7 September 2018. CF is very grateful to the organizers for a productive workshop and for financial support. TC was supported by ARC grant LP170100985.
\end{acknowledgements}


\newcommand{\etalchar}[1]{$^{#1}$}

\end{document}